\documentclass[a4paper,11pt]{article}
\pdfoutput=1
\usepackage[utf8]{inputenc}
\usepackage{jheppub}
\usepackage[T1]{fontenc}
\usepackage{amsmath}

\usepackage{graphicx}
\usepackage{color}
\usepackage{amsmath}
\usepackage{graphicx,textcomp,float,gensymb,wrapfig, enumitem,comment,dsfont,subfigure,framed,slashed,appendix,wrapfig,wasysym}
\usepackage{comment}
\usepackage[export]{adjustbox}
\usepackage[utf8]{inputenc} 

\newcommand{\gs}{g_\star}
\newcommand{\gss}{g_{\star \mathfrak{s}}}
\newcommand{\Trh}{T_\text{rh}}

\newcommand{\Tmax}{T_\text{max}}

\newcommand{\be}{\begin{equation}}
\newcommand{\ee}{\end{equation}}
\newcommand{\bea}{\begin{eqnarray}}  
\newcommand{\eea}{\end{eqnarray}}

\title{Kaluza-Klein FIMP Dark Matter\\in Warped Extra-Dimensions}

\author[a]{Nicolás Bernal,}
\author[b]{Andrea Donini,}
\author[b]{Miguel G. Folgado,}
\author[b]{Nuria Rius.}

\affiliation[a]{Centro de Investigaciones, Universidad Antonio Nariño,\\
Carrera 3 Este \# 47A-15, Bogotá, Colombia}
\affiliation[b]{Instituto de Física Corpuscular, Universidad de Valencia and CSIC,\\
Edificio Institutos Investigación, Catedrático Jose Beltrán 2, Paterna, 46980 Spain}

\emailAdd{nicolas.bernal@uan.edu.co}
\emailAdd{donini@ific.uv.es}
\emailAdd{\\migarfol@ific.uv.es}
\emailAdd{nuria.rius@ific.uv.es}

\abstract{We study for the first time the case in which Dark Matter (DM) is made of Feebly Interacting Massive Particles (FIMP) interacting just gravitationally with the standard model particles in an extra-dimensional Randall-Sundrum scenario. We assume that both the dark matter and the standard model are localized in the IR-brane and only interact via gravitational mediators, namely the graviton, the Kaluza-Klein gravitons and the radion.
We found that in the early Universe DM could be generated via two main processes: the direct freeze-in and the sequential freeze-in. 
The regions where the observed DM relic abundance  is produced are largely compatible with cosmological and collider bounds.
}

\begin{document}

\begin{flushright}
PI/UAN-2020-669FT\\
FTUV-20-0415.2168\\
IFIC/20-15
\end{flushright}

\maketitle

\section{Introduction}
The nature of Dark Matter (DM) and its interactions remain an open question in our effort to understand the Universe. 
Up to now, the only evidence about the existence of such dark component is via its gravitational effects. 
It could well be that DM has no other kind of interaction and, thus, it will be undetectable by current and future particle physics experiments. 
Moreover, in such a case the reheating temperature needs to be quite high (typically $\gtrsim 10^{16}$~GeV for DM mass of 10 TeV)
in order to generate the observed DM relic abundance via a purely gravitational interaction~\cite{Garny:2015sjg, Tang:2017hvq, Garny:2017kha, Bernal:2018qlk}, 
given the value of the Planck mass, $m_P  \sim 10^{19}$~GeV, which determines its strength.

This is true, however, only if we live in a four-dimensional space-time: in extra-dimensional scenarios, the gravitational interaction may be enhanced, 
either because the  fundamental Planck scale in $D$ dimensions is $m_D \ll m_P$ (as in the case of Large Extra Dimensions (LED)~\cite{Antoniadis:1990ew, Antoniadis:1997zg, ArkaniHamed:1998rs,Antoniadis:1998ig,ArkaniHamed:1998nn}),
or due to a warping of the space-time which induces an effective Planck scale $\Lambda$ in the four-dimensional brane such that $\Lambda \ll m_P$ 
(as in  Randall-Sundrum models (RS)~\cite{Randall:1999ee,Randall:1999vf}), or by a mixture of the two mechanisms (as it occurs in the more recent ClockWork/Linear Dilaton (CW/LD) model~\cite{Antoniadis:2011qw, Cox:2012ee,Giudice:2016yja, Giudice:2017fmj}).
As it is well known, this feature of the extra-dimensional scenarios has been  advocated as a solution to the so-called hierarchy problem, i.e., the huge hierarchy between the 
electroweak scale, $\Lambda_\text{EW} \sim $ 250~GeV, and  the Planck scale, which would generate corrections  of order of the Planck scale to the Higgs mass. 
These corrections would destabilize the electroweak scale unless either an enormous amount of fine-tuning is present or the Standard Model (SM) is the ultimate theory, 
which seems unlikely given the questions that are not explained within this framework (for instance, neutrino masses and baryogenesis, besides DM itself). 
In the extra-dimensional models mentioned above, the large hierarchy between the electroweak scale and the fundamental (or effective) Planck scale is eliminated, 
since the latter can be as low as ${\cal O}$(TeV).

As a consequence of such lower Planck scale in extra-dimensional models (either fundamental or effective), the gravitational interaction is enhanced, and a DM particle with just such interaction could become a WIMP, that is, a stable or cosmologically  long-lived weakly interactive massive particle, with mass typically in  the range 100 - 1000 GeV, and whose relic abundance is set via the freeze-out mechanism. 
This possibility has been thoroughly studied in the framework of the RS scenario~\cite{Lee:2013bua, Lee:2014caa, Han:2015cty, Rueter:2017nbk, Rizzo:2018ntg, Carrillo-Monteverde:2018phy, Rizzo:2018joy, Brax:2019koq, Folgado:2019sgz, Kumar:2019iqs} and
in a series of recent papers that study generic spin-2 mediators~\cite{Kang:2020huh,Chivukula:2020hvi,Kang:2020yul,Kang:2020afi}. 
It has also been considered in the context of the CW/LD model~\cite{Folgado:2019gie}.

In this work we again explore the RS framework for DM, yet analyzing a different  scenario  in which the relic abundance of DM is set via the so-called DM freeze-in production mechanism~\cite{McDonald:2001vt, Choi:2005vq, Kusenko:2006rh, Petraki:2007gq, Hall:2009bx} (for a recent review see Ref.~\cite{Bernal:2017kxu}).  In this case DM is a feebly interacting massive particle (FIMP), so that 
it never reaches thermal equilibrium with the SM thermal bath, and as a consequence its abundance remains smaller than the equilibrium one along the history of the Universe.
More specifically, here we focus on the sub-case of ultraviolet (UV) freeze-in~\cite{Elahi:2014fsa} for which the temperature of the thermal bath is always lower than the scale of new physics, 
which in our model is the effective Planck scale in the 4-dimensional brane, $\Lambda$, at which the gravitons become strongly interacting. 

In our setup we assume that both the SM and the DM particles are localized in the same 4-dimensional brane, and by definiteness we consider 
real scalar DM, only. We relax the request for the RS model to solve the hierarchy problem, and allow $\Lambda$ to vary in a wide range ($\Lambda \in [10^2, 10^{16}]$~GeV) to fully explore 
the parameter space that could lead to the correct DM relic abundance via freeze-in from a purely phenomenological perspective.
In order to have a consistent model, we stabilize the size of the extra-dimension by using the Goldberger-Wise mechanism \cite{Goldberger:1999uk}, which generates the required potential for
the four-dimensional radion field. Then, besides the interaction through Kaluza-Klein (KK) gravitons, we also take into account that the SM and DM species can interact with the radion.
We consider both SM particle annihilation into DM through KK-gravitons and the radion (direct freeze-in), as well as production of DM from out-of-equilibrium KK-gravitons and the radion (sequential freeze-in). 
We solve numerically the relevant Boltzmann equations in  all cases and also provide analytical approximations for the final DM relic abundance in different ranges of the temperature, 
useful to understand our main results. We always work within the sudden decay approximation for the inflaton, and shortly comment on how our findings would be affected by a non-instantaneous inflaton decay.
 
We vary the DM mass, radion and KK-graviton masses and the scale $\Lambda$, determining the reheating temperature  $\Trh$  which leads to the correct DM relic abundance in each case, within the validity range of our effective four-dimensional theory. We find that in this scenario the  observed DM density can be generated even with a reheating temperature lower than the electroweak scale. 
Recall that the only constraint on $\Trh$ is that it has to be higher than the Big Bang Nucleosynthesis temperature of around a few MeV~\cite{Sarkar:1995dd, Kawasaki:2000en, Hannestad:2004px, DeBernardis:2008zz, deSalas:2015glj, Hasegawa:2019jsa}.

The outline of the paper is as follows: in Sec.~\ref{sec:theo} we briefly remind the main features of the RS scenario; 
Sec.~\ref{sec:DMPEU} is devoted to the analysis of DM production via freeze-in within our model, both via direct and sequential freeze-in;
finally, in Sec.~\ref{sec:conclusions} we present our conclusions. Some details on the RS scenario are given in Apps.~\ref{app:KKdec} and \ref{app:radlag}, whereas 
the relevant interaction rates used in our calculations are collected in App.~\ref{app:annihil}.


\section{Theoretical Framework}
\label{sec:theo}

In this Section, we shortly remind some aspects of the Warped Extra-Dimension scenario (also called Randall-Sundrum model \cite{Randall:1999ee}) relevant in the rest of the paper. 
Some further details on RS scenarios are given in Apps.~\ref{app:KKdec} and \ref{app:radlag}.


The popular Randall-Sundrum scenario (from now on RS or RS1 \cite{Randall:1999ee}, to be distinguished from the scenario called RS2 \cite{Randall:1999vf})
consider a non-factorizable 5-dimensional metric in the form: 
\be
\label{eq:5dmetric}
ds^2 = e^{-2\sigma}\, \eta_{\mu \nu}\, dx^{\mu}dx^{\nu} - r_c^2 \, dy^2\,,
\ee
where $\sigma = k\, r_c\, |y|$ and the signature of the metric is $(+,-,-,-,-)$.
In this scenario, $k$ is the curvature along the 5$^\text{th}$-dimension and it is ${\cal O} \left ( M_P \right )$. 
The length-scale $r_c$, on the other hand, is related to the size of the extra-dimension: we only consider a slice of the space-time between two branes
located conventionally at the two fixed-points of an orbifold, $y = 0$ (the so-called UV-brane) 
and $y = \pi$ (the IR-brane). The 5-dimensional space-time is a slice of ${\cal }AdS_5$ and 
the exponential factor that multiplies the ${\cal M}_4$ Minkowski 4-dimensional space-time 
is called the ``warp factor''. 

The action in 5D is: 
\begin{equation}
S = S_{\rm gravity} + S_{\rm IR} + S_{\rm UV}
\end{equation}
where
\begin{equation}
\label{eq:5dgravity}
S_{\rm gravity} = \frac{16 \pi}{M_5^3} \int d^4 x \, \int_0^\pi r_c\, dy\, \sqrt{G^{(5)}} \, \left [ R^{(5)}  
- 2 \Lambda_5  \right ] \, ,
\end{equation}
with $M_5$ the fundamental gravitational scale, $G^{(5)}$ and $R^{(5)}$ the 5-dimensional metric and Ricci scalar, respectively, and $\Lambda_5$ the 5-dimensional 
cosmological constant. 
As usual, we consider capital Latin indices $M$, $N$ to run over the 5 dimensions and Greek indices $\mu$, $\nu$ only over 4 dimensions.
The reduced Planck mass  is related to the fundamental scale $M_5$ as: 
\be
M_P^2 = \frac{M_5^3}{k} \,  \left (1-e^{-2k\, \pi\, r_c} \right )  \, ,
\ee
where $M_P = m_P/\sqrt{8\pi} \simeq 2.435 \times 10^{18}$~GeV, being $m_P$ the Planck mass.

We consider for the two brane actions  the following expressions: 
\begin{equation}
	S_{\rm IR} = \int d^4 x \sqrt{- g^{(4)}} \, \left[ - f_{\rm IR}^4 + {\cal L}_{\rm SM} + {\cal L}_{\rm DM}  \right]
\end{equation}
and
\begin{equation}
\label{eq:UVlagr}
	S_{\rm UV} = \int d^4 x \sqrt{-g^{(4)}} \, \left[ - f_{\rm UV}^4 + \dots \right] ,
\end{equation}
where $f_{\rm IR}$, $f_{\rm UV}$ are the brane tensions for the two branes, ${\cal L}_\text{SM}$ and ${\cal L}_\text{DM}$ the SM and DM Lagrangians densities, respectively.
Notice that in 4-dimensions in general $\eta_{\mu\nu}$ is replaced by $g^{(4)}_{\mu\nu}$,  the 4-dimensional induced metric on the brane.
Dots in eq.~(\ref{eq:UVlagr}) stand for any possible new physics on the UV brane and, thus, decoupled from us.

In RS scenarios, in order to achieve the metric in eq.~(\ref{eq:5dmetric}) as a classical solution of the Einstein equations, the brane-tension terms in $S_{\rm UV}$ and $S_{\rm IR}$ are chosen such as to cancel the 5-dimensional cosmological constant, 
$f_{\rm IR}^4 = - f_{\rm UV}^4 = \sqrt{- 24 M_5^3 \, \Lambda_5}$.
Throughout this paper, we consider all the SM and DM fields localized on the IR-brane, whereas on the UV-brane we could have any other physics that is Planck-suppressed. We assume that DM particles only interact with the SM particles gravitationally.\footnote{
If the DM particle is a scalar singlet under the SM gauge group, it will also interact with the SM through its mixing with the Higgs boson.}

Alternative DM spectra (with particles of spin higher than zero or with several 
particles) will not be studied here. Notice that, in 4-dimensions, the gravitational interactions would be enormously suppressed by powers of the Planck mass. However, in an extra-dimensional scenario, the gravitational interaction is actually enhanced: on the IR--brane, in fact, the effective gravitational coupling is 
$\Lambda = M_P\, \exp\left(- k\, \pi\, r_c\right)$, due to the rescaling factor $\sqrt{G^{(5)}}/\sqrt{- g^{(4)}}$.
It is easy to see that $\Lambda \ll M_P$ even for moderate choices of $\sigma$. In particular, for 
$\sigma = k\,  r_c \simeq 10$ the RS scenario can address the hierarchy problem.
From a purely phenomenological perspective, here we will work with $\Lambda = [10^2,\,10^{16}]$~GeV, relaxing the requirement that the RS model should provide a
solution to the hierarchy problem. 
 
The Kaluza-Klein decomposition of 5-dimensional fields in a RS scenario is shortly reviewed in App.~\ref{app:KKdec}. The coupling between KK-gravitons 
and brane matter (being $h_{MN}$ the 5D graviton field and $h_{\mu\nu}$ its 4D component) is:
\bea
\mathcal{L} &=& -\frac{1}{M_5^{3/2}} T^{\mu \nu}(x) h_{\mu \nu}(x,y=\pi) = -\frac{1}{M_5^{3/2}} T^{\mu \nu}(x) \sum_{n=0} h^{n}_{\mu \nu}\frac{\chi^{n}}{\sqrt{r_c}} \, , \nonumber \\
&=& -\frac{1}{M_P} T^{\mu \nu}(x) h^{0}_{\mu \nu}(x) -\frac{1}{\Lambda} \sum_{n=1} T^{\mu \nu}(x) h^{n}_{\mu \nu}(x) \, ,
\eea
from which is clear that the coupling between KK-graviton modes with $n \neq 0$ is suppressed by the effective scale $\Lambda$ and not by the Planck scale. 


Stabilizing the size of the extra-dimension to be $y = \pi\, r_c$ is not easy. Long ago it was shown 
that bosonic quantum loops have a net effect on the border of the extra-dimension such that the extra-dimension itself should shrink to a point~\cite{Appelquist:1982zs,Appelquist:1983vs,deWit:1988xki}.
This feature, in a flat extra-dimension, can only be compensated by fermionic quantum loops
and, usually, some supersymmetric framework is invoked to stabilize the radius of the extra-dimension (see, e.g., Ref.~\cite{Ponton:2001hq}). A popular mechanism implemented in RS 
models to stabilize the size of the extra-dimension was proposed in Refs.~\cite{Goldberger:1999wh,Goldberger:1999uk} and can be summarized as follows: if we add a bulk scalar field $S$
with a scalar potential $V(S)$ and some {\em ad hoc}  localized potential terms, $\delta (y=0) V_{\rm UV}(S)$ and $\delta (y = \pi\, r_c) V_{\rm IR} (S)$, it is possible to generate an effective potential $V(\varphi)$ for the four-dimensional field $\varphi = f \, \exp \left ( - k\, \pi\, T \right )$
(with $f = \sqrt{24 M_5^3/k}$ and $\langle T \rangle  = r_c$). 
The minimum of this potential can yield the desired value of $k r_c$ without extreme fine-tuning of the parameters.

The $S$ field will generically mix with the graviscalar $G^{(5)}_{55}$ (notice that the KK-tower of the graviscalar is absent from the low-energy spectrum, as they are eaten by the KK-tower 
of graviphotons to get a mass due to the spontaneous breaking of translational invariance caused by the presence of one or more branes). On the other hand, the KK-tower of the field $S$ is present, but heavy (see Ref.~\cite{Goldberger:1999un}). The only light field present in the spectrum is, then, a combination of the graviscalar zero-mode and the $S$ zero-mode.
This field is usually called the {\em radion}, $r$. Its mass can be obtained from the effective potential 
$V (\varphi)$ and is given by $m_\varphi^2 = k^2 v_v^2 / 3 M_5^3 \, \epsilon^2 \, \exp (-2 \pi\, k\, r_c)$, where $v_v$ is the value of $S$ at the visible brane and $\epsilon = m^2/4 k^2$ (with $m$ 
the mass of the field $S$). Quite generally $\epsilon \ll 1$ and, therefore, the mass of the radion can be much smaller than the first KK-graviton mass.

The radion, as for the KK-graviton case, interacts with both the DM and SM particles. It couples with matter through the trace of the energy-momentum tensor $T$ \cite{Lee:2013bua}. Massless gauge fields do not contribute to the trace of the energy-momentum tensor,  but effective couplings are generated from two different sources: quarks and $W$ boson loops  and the trace anomaly 
\cite{Blum:2014jca}. Thus the radion  Lagrangian takes the following form \cite{Goldberger:1999un, Csaki:1999mp}:
\be
\mathcal{L}_r = \frac{1}{2}(\partial_\mu r)(\partial^\mu r) - \frac{1}{2} m_r^2 r^2 +  \frac{1}{\sqrt{6}\Lambda} r T+ \frac{\alpha_\text{EM} \, C_\text{EM}}{8\pi\sqrt{6}\Lambda} r F_{\mu\nu} F^{\mu\nu} + \frac{\alpha_{S}C_{3}}{8 \pi \sqrt{6} \Lambda} r \sum_a F^a_{\mu\nu} F^{a\mu\nu}\, ,
\ee
where $F_{\mu\nu}$, $F^a_{\mu\nu}$ are the Maxwell and $SU(3)_c$ Yang-Mills tensors, respectively.
Further details on the radion lagrangian can be found in App.~\ref{app:radlag}.

Possible couplings between KK-modes of the bulk scalar field $S$, the DM and SM fields are usually allowed, in the absence of some {\em ad hoc} bulk symmetry to forbid them. In the rest of 
the paper we will not include them, 
since we want to focus on just gravitational mediators (radion and KK-gravitons) between the SM and the dark particles.

Finally, we want to comment about the AdS/CFT correspondence, which suggests a duality between strongly coupled conformal  field theories in 4D and weakly coupled gravity in 5D (see, for example, Ref.~\cite{Aharony:1999ti} and refs. therein), also called holography. Within this framework, the extra-dimensional model described above can be interpreted as a strongly interacting theory in which the particles  localized at the IR-brane are bound states, while the presence of  gravity mediators (KK-gravitons and radion) is a consequence of the conformal symmetry of the composite sector, spontaneously broken by the strong dynamics. The radion is thought to be the Goldstone boson of dilatation symmetry in 4D, i.e., the dilaton, although the dual interpretation of the massive gravitons is not so well understood \cite{Lee:2013bua}. 
The scale $\Lambda$ in the holographic dual corresponds to the scale of conformal symmetry breaking in 4D.

\section{Dark Matter Production in the Early Universe}
\label{sec:DMPEU}

In Refs.~\cite{Folgado:2019sgz,Folgado:2019gie} some of us have studied how to reach
the observed DM relic abundance in the freeze-out scenario. Freeze-out occurs
if the interactions between DM and SM particles are strong enough
to bring them into chemical equilibrium. However, if the interaction rates between the visible and the dark sectors were never strong enough, the observed DM relic abundance could still have been produced in the early Universe by non-thermal processes. This is what occurs in the so-called
freeze-in mechanism.

The evolution of the DM, radion and KK-gravitons number densities ($n$, $n_r$ and $n_K$ respectively) is given by a
 system of coupled Boltzmann equations:
\begin{eqnarray} \label{eq:cosmo1B}
	\frac{dn}{dt}+3\,H\,n&=&-\gamma_\text{DM$\to$SM}\left[\left(\frac{n}{n^\text{eq}}\right)^2-1\right]+\gamma^d_\text{KK$\to$DM}\left[\frac{n_K}{n_K^\text{eq}}-\left(\frac{n}{n^\text{eq}}\right)^2\right],\\
	\frac{dn_r}{dt}+3\,H\,n_r&=&-\gamma_\text{r$\to$SM}\left[\left(\frac{n_r}{n_r^\text{eq}}\right)^2-1\right]-\gamma^d_\text{r$\to$DM}\left[\frac{n_r}{n_r^\text{eq}}-\left(\frac{n}{n^\text{eq}}\right)^2\right]\nonumber\\
	&&-\gamma^d_\text{r$\to$SM}\left[\frac{n_r}{n_r^\text{eq}}-1\right],\\ 
	\frac{dn_K}{dt}+3\,H\,n_K&=&-\gamma_\text{KK$\to$SM}\left[\left(\frac{n_K}{n_K^\text{eq}}\right)^2-1\right]-\gamma^d_\text{KK$\to$DM}\left[\frac{n_K}{n_K^\text{eq}}-\left(\frac{n}{n^\text{eq}}\right)^2\right]\nonumber\\
	&&-\gamma^d_\text{KK$\to$SM}\left[\frac{n_K}{n_K^\text{eq}}-1\right], \label{eq:cosmo2B}
\end{eqnarray}
where $H$ corresponds to the Hubble expansion rate, and $n_i^\text{eq}$ are the number densities at equilibrium of the species $i$. 
Interactions that {\it only} involve bulk particles, namely KK-gravitons and radions, both in the initial and final states are subdominant due to a strong suppression of $1/\Lambda^8$.
The quantity $\gamma_\text{$\Phi\to$SM}$ is the interaction rate density for the 2-to-2 annihilations of a field $\Phi$ (either DM, KK-graviton or radion) into SM particles.
Similarly, $\gamma^d_\text{$\Phi\to$DM}$ and $\gamma^d_\text{$\Phi\to$SM}$ are the interaction rate densities for the 2-body decay of a field $\Phi$ into DM and SM particles, respectively.
Let us notice that in this extra-dimensional picture we need a Boltzmann equation like eq.~\eqref{eq:cosmo2} for every KK-mode.

A standard way to rewrite  the Boltzmann equations is using the dimensionless yield $Y\equiv n/\mathfrak{s}$, 
with $\mathfrak{s}$ the SM entropy density (not to be confused with the Mandelstam variable  $s$). 
The SM entropy density is defined, as  a function of the temperature, as $\mathfrak{s}(T)=\frac{2\pi^2}{45}\,\gss(T)\,T^3$ (where  $\gss(T)$ 
is the effective number of relativistic degrees of freedom~\cite{Drees:2015exa}).
Equations~\eqref{eq:cosmo1B} to~\eqref{eq:cosmo2B} can therefore be rewritten as
\begin{eqnarray}\label{eq:cosmo1}
	\frac{dY}{dT}&=&-\frac{\gamma_\text{DM$\to$SM}}{H\,\mathfrak{s}\,T}\left[\left(\frac{Y}{Y^\text{eq}}\right)^2-1\right]+\frac{\gamma^d_\text{KK$\to$DM}}{H\,\mathfrak{s}\,T}\left[\frac{Y_K}{Y_K^\text{eq}}-\left(\frac{Y}{Y^\text{eq}}\right)^2\right],\\
	\frac{dY_r}{dT}&=&-\frac{\gamma_\text{r$\to$SM}}{H\,\mathfrak{s}\,T}\left[\left(\frac{Y_r}{Y_r^\text{eq}}\right)^2-1\right]-\frac{\gamma^d_\text{r$\to$DM}}{H\,\mathfrak{s}\,T}\left[\frac{Y_r}{Y_r^\text{eq}}-\left(\frac{Y}{Y^\text{eq}}\right)^2\right]-\frac{\gamma^d_\text{r$\to$SM}}{H\,\mathfrak{s}\,T}\left[\frac{Y_r}{Y_r^\text{eq}}-1\right],\quad\label{eq:cosmo3}\\
	\frac{dY_K}{dT}&=&-\frac{\gamma_\text{KK$\to$SM}}{H\,\mathfrak{s}\,T}\left[\left(\frac{Y_K}{Y_K^\text{eq}}\right)^2-1\right]-\frac{\gamma^d_\text{KK$\to$DM}}{H\,\mathfrak{s}\,T}\left[\frac{Y_K}{Y_K^\text{eq}}-\left(\frac{Y}{Y^\text{eq}}\right)^2\right]\nonumber\\
	&&-\frac{\gamma^d_\text{KK$\to$SM}}{H\,\mathfrak{s}\,T}\left[\frac{Y_K}{Y_K^\text{eq}}-1\right].\label{eq:cosmo2}
\end{eqnarray}

In the freeze-in paradigm DM never gets in thermal equilibrium with the rest of the SM particles of the primordial plasma. It is usually assumed that after inflation the abundance of DM was negligible, and slowly produced via interaction between the SM particles.
Along the evolution of the Universe, the DM abundance was generated via two main processes:
\begin{enumerate}
	\item \underline{Direct freeze-in.}
		The DM abundance is generated directly by the annihilation of SM particles via an 
		$s$-channel exchange of KK-gravitons or a radion.
	\item \underline{Sequential freeze-in or freeze-in from the dark sector.}
		The DM abundance is generated by decays of KK-gravitons or radions, previously 
		produced by annihilations or inverse decays of SM particles via direct freeze-in.
		This scenario has been doubted ``sequential freeze-in''~\cite{Hambye:2019dwd}.
\end{enumerate}
Another production channel corresponds to the case in which the DM abundance is set entirely in the hidden sector by 4-to-2 interactions~\cite{Bernal:2015xba,Bernal:2017mqb,Bernal:2020gzm}.
However, such a possibility is sub-dominant due to a strong suppression by higher orders of the scale $\Lambda$.
It has been also shown that, independently of the nature of DM, it is possible to populate the relic abundance through a freeze-in mechanism via the exchange 
of a massless spin-2 graviton~\cite{Garny:2015sjg, Tang:2017hvq, Garny:2017kha, Bernal:2018qlk}. However, for this mechanism to be dominant, 
reheating temperatures $\Trh$ of the order of $10^{13}$~GeV for a DM mass of 1~MeV are required. We will see in the following that, in this warped extra-dimensional setup
(with KK-gravitons and the radion as additional fields playing the freeze-in mechanism) a much wider range of $\Trh$ is indeed possible.

These two main mechanisms previously mentioned, i.e. the direct and the sequential freeze-in, will be described in detail in the following subsections.

\subsection{Direct Freeze-in}
\label{sec:directFI}

As it was briefly sketched above, in the case of direct freeze-in the DM abundance $n$ is generated by the annihilation of SM particles via an $s$-channel exchange of KK-gravitons or a radion.%
\footnote{Another possibility corresponds to the interactions mediated by Higgs bosons. However, we focus here on the extra-dimensional portal ignoring the Higgs one.
This can be reached by assuming a quartic coupling $\lambda_{h\chi}$ between the Higgs and the DM such as $\lambda_{h\chi}\ll 10^{-10}$~\cite{Yaguna:2011qn, Bernal:2018kcw}.}
If the production cross-section is small enough to keep DM out of chemical equilibrium with the SM bath, and the evolution of the DM abundance $n$ (or of the yield $Y$) 
is largely dominated by the interaction rate density $\gamma_\text{DM$\to$SM}$, eqs.~\eqref{eq:cosmo1} to~\eqref{eq:cosmo2} can be simplified to:
\begin{equation}\label{eq:BEapproxDFI0}
	\frac{dY}{dT}\simeq\frac{\gamma_\text{DM$\to$SM}}{H\,\mathfrak{s}\,T}\left[\left(\frac{Y}{Y^\text{eq}}\right)^2-1\right]\simeq-\frac{\gamma_\text{DM$\to$SM}}{H\,\mathfrak{s}\,T}\,.
\end{equation}
In a Universe dominated by SM radiation the Hubble expansion rate is $H^2=\frac{\rho_\text{SM}}{3\,M_P^2}$, where the SM energy density is $\rho_\text{SM}(T)=\frac{\pi^2}{30}\gs(T)\,T^4$ and $\gs(T)$ is the effective numbers of relativistic degrees of freedom for the SM radiation~\cite{Drees:2015exa}. Then, eq.~\eqref{eq:BEapproxDFI0} becomes:
\begin{equation}\label{eq:BEapproxDFI}
	Y(T)\simeq\frac{135}{2\pi^3\,\gss}\sqrt{\frac{10}{\gs}}\,M_P\int_{\Trh}^T\frac{\gamma_\text{DM$\to$SM} (T)}{T^6}\,dT\,,
\end{equation}
where $T_{\rm rh}$ is the reheating temperature which, in the approximation of a sudden decay of the inflaton, corresponds to the maximal temperature reached by the SM thermal bath.
In order to get eq.~(\ref{eq:BEapproxDFI}) a vanishing initial DM abundance at $T=\Trh$ was assumed and the temperature dependence of $\gs(T)$ and $\gss(T)$ has been neglected.
The asymptotic values $\gs$ and $\gss$ correspond to the SM values for $T \gg m_t$, $\gs = \gss = 106.75$ (which take into account all SM degrees of freedom). 
Since this approximation is reliable for temperatures above the QCD phase transition, we explore the range $\Trh \gtrsim 1$~GeV.

\begin{figure}[t]
	\begin{center}
		\includegraphics[height=0.32\textwidth]{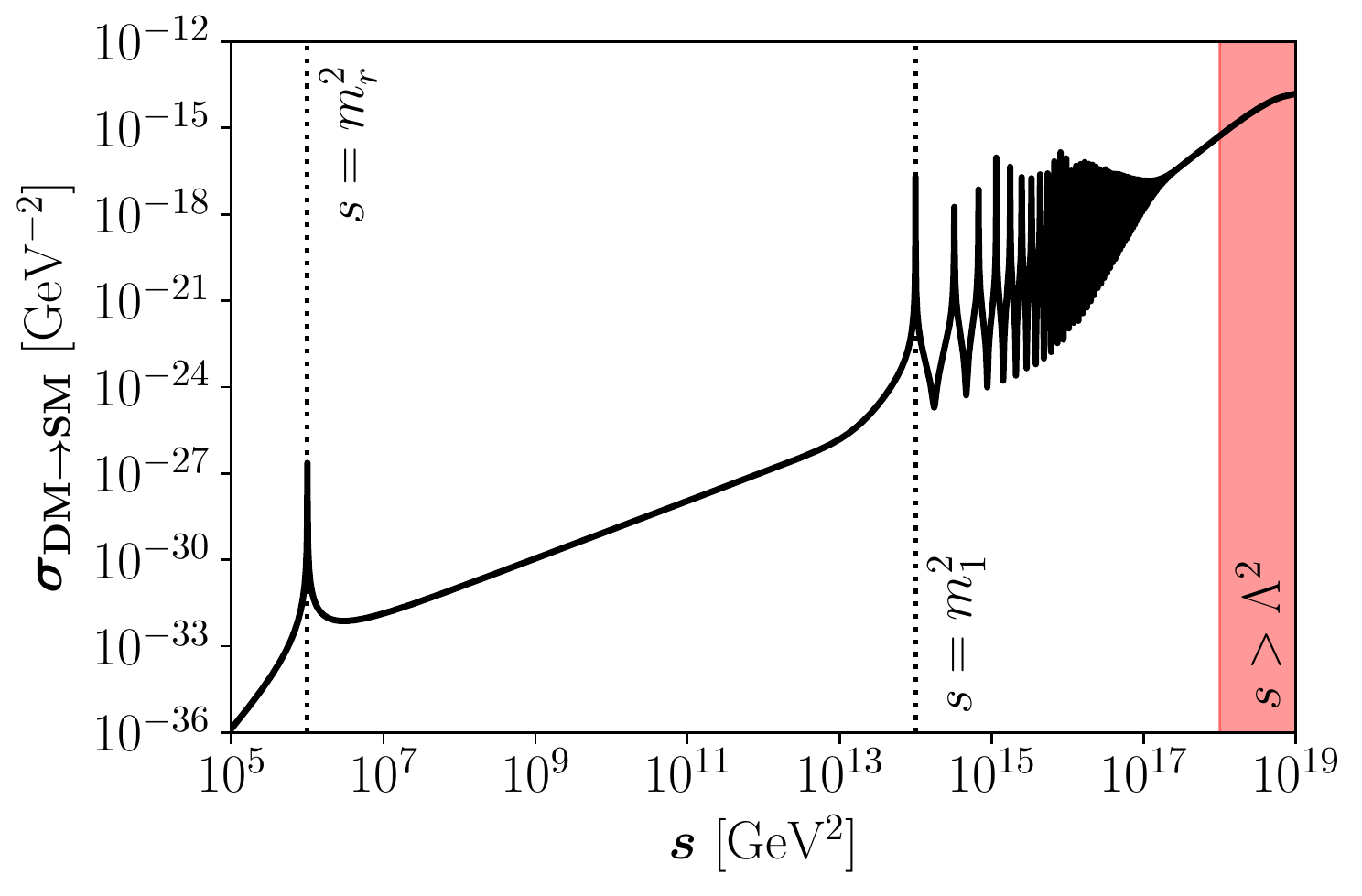}
		\includegraphics[height=0.32\textwidth]{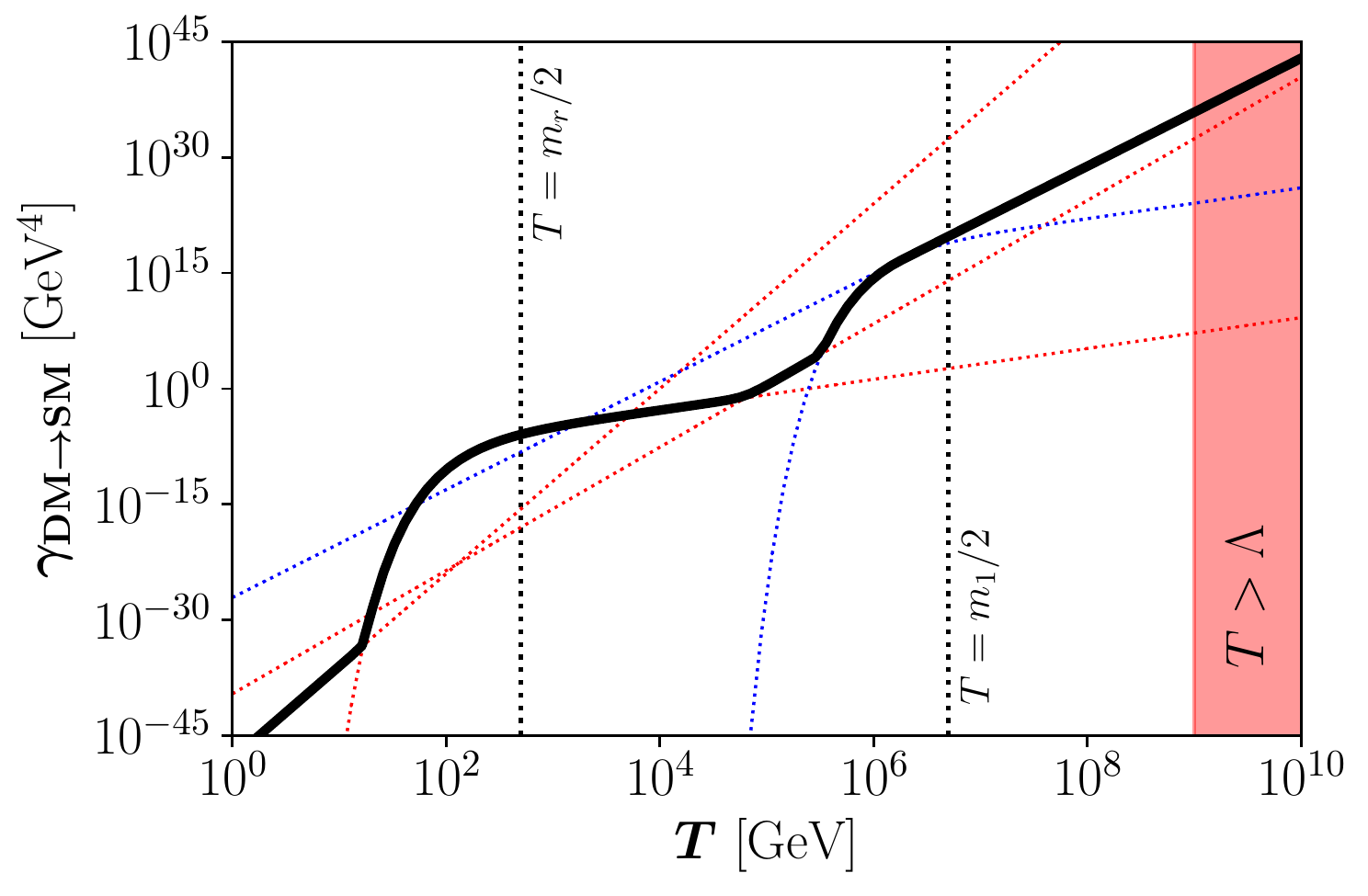}
		\caption{Black solid lines represent the DM annihilation cross section (left panel) and interaction rate density (right panel) for $m_r=10^3$~GeV,  $m_1=10^7$~GeV and $\Lambda=10^9$~GeV.
	Colored lines depict the analytical approximations of eq.~\eqref{eq:gamma}, where red and blue stand for interactions dominated by the exchange of a radion or a KK-gravitons, respectively.
	The red-shaded regions on the right of both panels are beyond our EFT approach.}\label{fig:cross}
	\end{center}
\end{figure}

The interaction rate density $\gamma_\text{DM$\to$SM}$ can be computed from the total DM annihilation cross-section into SM states $\sigma_\text{DM$\to$SM}$ which,
in the limit where the DM and SM particle masses are negligible, can be expressed as:%
\footnote{The details of the individual cross-sections are reported in Appendix~\ref{app:DMannihila}.}
\begin{equation}
\label{virtual_exchange}
	\sigma_\text{DM$\to$SM}(s) \simeq \frac{49}{1440 \pi} \frac{s^3}{\Lambda^4}\,\left|\sum_{n=1}^{\infty} \frac{1}{s-m_n^2 + i\,m_n\,\Gamma_n}\right|^2 + \frac{s^3}{288 \pi \Lambda^4} \frac{1}{(s - m_r^2)^2 + m_r^2\,\Gamma_r^2}\,,
\end{equation}
where the two terms correspond to the exchange of KK-gravitons and the radion, respectively.
Left panel of Fig.~\ref{fig:cross} shows with a solid black line an example of the DM annihilation cross section $\sigma_\text{DM$\to$SM}$ 
for a particular point in the parameter space, $m_r=10^3$~GeV, $m_1=10^7$~GeV and $\Lambda=10^9$~GeV.
Notice that this cross-section is largely independent of the DM mass, $m_\chi$, as long as $m_\chi^2 \ll s$. Then, also the interaction rate density becomes independent 
of $m_\chi$ provided $m_\chi \ll T$. Therefore, in the following we will consider as a benchmark point $m_\chi = 1$ MeV to illustrate our results, but keeping in mind that 
they can be extended  to a wide range of DM masses, typically between the keV and PeV scale.
The first peak at $s=m_r^2$ corresponds to the resonant exchange of a radion, whereas the following well-separated peaks correspond to the lightest KK-graviton modes. The 
non-trivial behavior for $s\gg m_1^2$ is due to the sum over poles and interferences of many different KK mediators.
For very large values of the KK-number $n$, the widths of the KK-graviton resonances become comparable to their mass gap, $\Gamma_n(\sqrt{s}) \simeq \Delta m$. This happens
approximately for:
\be
s \gtrsim \Lambda^{4/3} \left( \frac{240 \, \pi^2 \, m_1 }{73 \, x_1} \right)^{2/3} \, ,
\ee
as at large $n$ the KK-modes separation is a constant, $\Delta m \simeq m_1/x_1$, see eq.~(\ref{eq:KKmasseslargen}). 
In this regime the resonances overlap and become individually indistinguishable. They eventually merge
into one single contribution to the cross-section, as it can be seen in the rightmost region of Fig.~\ref{fig:cross} (left).
Finally, the red-shaded region corresponding to $s>\Lambda^2$ is beyond our EFT approach, being the center-of-mass energy of the process larger than the effective scale of the theory.

In order to solve eq.~(\ref{eq:BEapproxDFI}), we need to compute the interaction rate density $\gamma_\text{DM$\to$SM}$ as a function of the temperature. 
In general, for the process where two particles ($i$, $j$) annihilate into two states ($k$, $l$), the interaction rate density $i + j \rightarrow  k + l$ is defined as:
\begin{equation}\label{eq:gamma_sin_approx}
        \gamma(T)=\frac{T}{64\,\pi^4} \int_{s_\text{min}}^{\infty}ds \sqrt{s}\,\sigma_R(s)\,K_1\left(\frac{\sqrt{s}}{T}\right)\,,
\end{equation}
where $s_{\text{min}}\equiv \text{max}\left[(m_i + m_j)^2,\, (m_k + m_l)^2\right]$, $\sigma_R$ is the reduced cross-section summed over all the degrees of freedom of the initial 
and final states, and $K_1$ is the modified Bessel function.
$\sigma_R$ corresponds to the total cross-section $\sigma(s)$ without the flux factor, and can be written as:
\be
\sigma_{R}(s) = 2\,\frac{\left[s-(m_i+m_j)^2\right]\left[s-(m_i-m_j)^2\right]}{s}\,\sigma(s)\,.
\ee

Several useful approximations can be implemented for different ranges of $T$, such that the interaction rate density $\gamma_\text{DM$\to$SM}$ for the DM annihilation into SM states 
becomes:
\begin{equation}\label{eq:gamma}
	\gamma_\text{DM$\to$SM}(T)\simeq
	\begin{cases}
		\left ( \frac{1}{\Lambda^4\,m_r^4} \right ) \,  T^{12} \qquad&\text{for }T\ll\frac{m_r}{2},\\[8pt]
		10^{-6}\, \left ( \frac{m_r^8 }{\Lambda^4\, \Gamma_r} \right ) \, T \, K_1(\frac{m_r}{T})\qquad&\text{for }T \simeq \frac{m_r}{2},\\[8pt]
		3\times 10^{-4}\, \left ( \frac{1}{\Lambda^4} \right ) \, T^{8} \qquad&\text{for }\frac{m_r}{2}\ll T \ll\frac{m_1}{2},\\[8pt]
		10^{-5}\,\left ( \frac{m_1^8}{\Lambda^4\,\Gamma_1} \right ) \,  T \, K_1\left(\frac{m_1}{T}\right)\qquad&\text{for }T\simeq\frac{m_1}{2},\\[8pt]
		7\times 10^{-4}\, \left ( \frac{m_1^2}{\Lambda^4\,\Gamma_1} \right ) \, T^7 \qquad&\text{for } T\gg \frac{m_1}{2}.
	\end{cases}\\
\end{equation}
The right panel of Fig.~\ref{fig:cross} shows the DM interaction rate density for $m_r=10^3$~GeV, $m_1=10^7$~GeV and $\Lambda=10^9$~GeV with a black solid line, whose 
behavior as a function of the temperature can be easily understood using the approximations in eq.~(\ref{eq:gamma}):
\begin{itemize}
	\item At low temperatures ($T\ll m_r/2$) all the mediators are very heavy and  decouple from the low-energy theory; the rate presents a strong temperature dependence, $\gamma\propto T^{12}$, represented by a red-dotted straight line in the plot.
	\item When $T\simeq m_r/2$, the resonant exchange of a radion dominates and $\gamma\propto T\,K_1(m_r/T)$. This corresponds to the first bump in the plot, again coinciding with a red-dotted (curved) line.
	\item In the intermediate regime, $m_r/2 \ll T\ll m_1/2$, the temperature is higher than the radion mass but still smaller than all KK states. The interaction is, thus, driven by the exchange of the light radion, with $\gamma\propto T^8$. This is shown by the second straight red-dotted line in the plot, with a slope smaller than the first one (as it is proportional to $T^8$, compared to $T^{12}$ in the first region).
	\item We reach then the region in which the KK-gravitons dominance takes over: first, at the peak of the first KK-graviton mode ($T \simeq m_1/2$) for which $\gamma\propto T\,K_1(m_1/T)$, corresponding to the second bump in the plot.
	\item Eventually, when the increase of the temperature makes heavier KK-graviton states to have a sizable contributions to the rate, with a constructive interference giving a $\gamma\propto T^7$ behavior.
\end{itemize}
We can see that all the different regimes  in $T$ follow extremely well the curved and straight blue- and red-dotted lines, corresponding to the approximate behaviors depicted in 
eq.~(\ref{eq:gamma}).
As for the left panel, the red-shaded region corresponding to $T>\Lambda$ is beyond our EFT approach.

Notice that a big hierarchy between $m_r$ and $m_1$ was chosen in order to avoid an overlap between the two bumps, such that the five regimes in eq.~\eqref{eq:gamma} can 
be clearly seen in the plot. For generic choices in the parameter space, overlap between regions may occur.

Using the approximated expressions of $\gamma_\text{DM$\to$SM}$ from eq.~\eqref{eq:gamma},  the Boltzmann equation \eqref{eq:BEapproxDFI} 
can be analytically solved, finding for the different regions in $T$:
\begin{equation}\label{eq:yield_direct}
	Y_0\simeq
	\begin{cases}
		\frac{3\times 10^{-1}}{\gss}\sqrt{\frac{10}{\gs}} \left (\frac{M_P}{m_r^4\,\Lambda^4} \right ) \,
		                                              \Trh^7\quad&\text{for }\Trh\ll m_r/2\,,\\[8pt]
		\frac{6.7\times 10^{-7}}{\gss}\sqrt{\frac{10}{\gs}} \left ( \frac{M_P\,m_r^{9/2}}{\Lambda^4\Gamma_r} \right ) \, 
		\left ( \frac{4m_r^2+10m_r\,\Trh+15\Trh^2}{\Trh^{5/2}} \right ) \, e^{-\frac{m_r}{\Trh}}\quad&\text{for }\Trh\simeq m_r/2\,,\\[8pt]
		\frac{2\times 10^{-4}}{\gss}\sqrt{\frac{10}{\gs}} \left ( \frac{M_P}{\Lambda^4} \right ) \, \Trh^3\quad&\text{for }m_r/2\ll\Trh\ll m_1/2\,,\\[8pt]
		\frac{6.7\times 10^{-6}}{\gss}\sqrt{\frac{10}{\gs}} \left ( \frac{M_P\,m_1^{9/2}}{\Lambda^4\Gamma_1} \right ) \, \left ( \frac{4m_1^2+10m_1\,\Trh+15\Trh^2}{\Trh^{5/2}} \right ) 
		\,e^{-\frac{m_1}{\Trh}}\quad&\text{for }\Trh\simeq m_1/2\,,\\[8pt]
		\frac{8\times 10^{-4}}{\gss}\sqrt{\frac{10}{\gs}} \left ( \frac{M_P \, m_1^2}{\Lambda^4 \, \Gamma_1} \right ) \,\Trh^2\quad&\text{for }\Trh \gg m_1/2 \,, \\[8pt]
	\end{cases}
\end{equation}
where $Y_0$ corresponds to the asymptotic value of $Y(T)$ for $T\ll\Trh$.
The final DM yield in eq.~\eqref{eq:yield_direct} has a strong dependence on $\Trh$, characteristic of the UV freeze-in production mechanism.

Finally, let us emphasize that for the previous analysis to be valid, the DM has to be out of chemical equilibrium with the SM bath.
One needs to guarantee, therefore, that the interaction rate density  be $\gamma_\text{DM$\to$SM}\ll n^\text{eq}\,H$, which translates into:
\begin{equation}
	\Trh\ll
	\begin{cases}
		0.7\left(\frac{\gs}{10}\right)^{1/14}\left(\frac{\Lambda^4 \, m_r^4}{M_P}\right)^{1/7} \, \qquad&\text{for }\Trh\ll m_r/2\,,\\[8pt]
		-\frac27m_1/W_{-1}\left[-7.8\left(\sqrt{\frac{\gs}{10}}\frac{\Lambda^4\,\Gamma_r}{m_r^4\,M_P}\right)^{2/7}\right]\qquad&\text{for }\Trh\simeq m_r/2\,,\\[8pt]
		7.5\left(\frac{\gs}{10}\right)^{1/6}\left(\frac{\Lambda^4}{M_P}\right)^{1/3} \, \qquad&\text{for }m_r/2\ll\Trh\ll m_1/2\,,\\[8pt]
		-\frac27m_1/W_{-1}\left[-4.3\left(\sqrt{\frac{\gs}{10}}\frac{\Lambda^4\,\Gamma_1}{m_1^4\,M_P}\right)^{2/7}\right]\qquad&\text{for }\Trh\simeq m_1/2\,,\\[8pt]
		13.5\left(\frac{\gs}{10}\right)^{1/4}\sqrt{\frac{\Gamma_1}{M_P}}\,\frac{\Lambda^2}{m_1}\qquad&\text{for }\Trh \gg m_1/2 \,,
	\end{cases}
\end{equation}
where $W_{-1} [x]$ corresponds to the $-1$ branch of the Lambert $W$ function computed at $x$.

\begin{figure}[t]
\begin{center}
\includegraphics[height=0.33\textwidth]{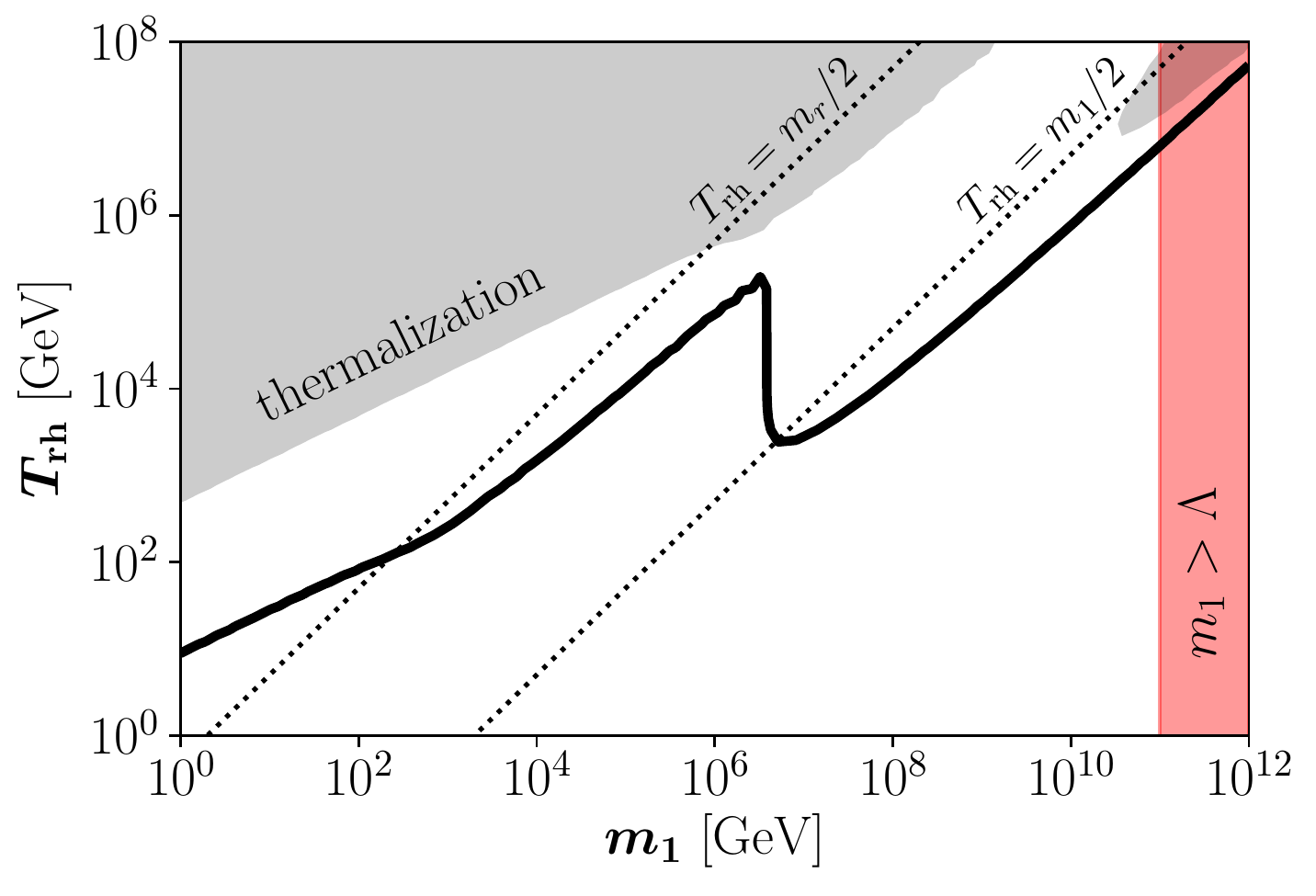}
\includegraphics[height=0.33\textwidth]{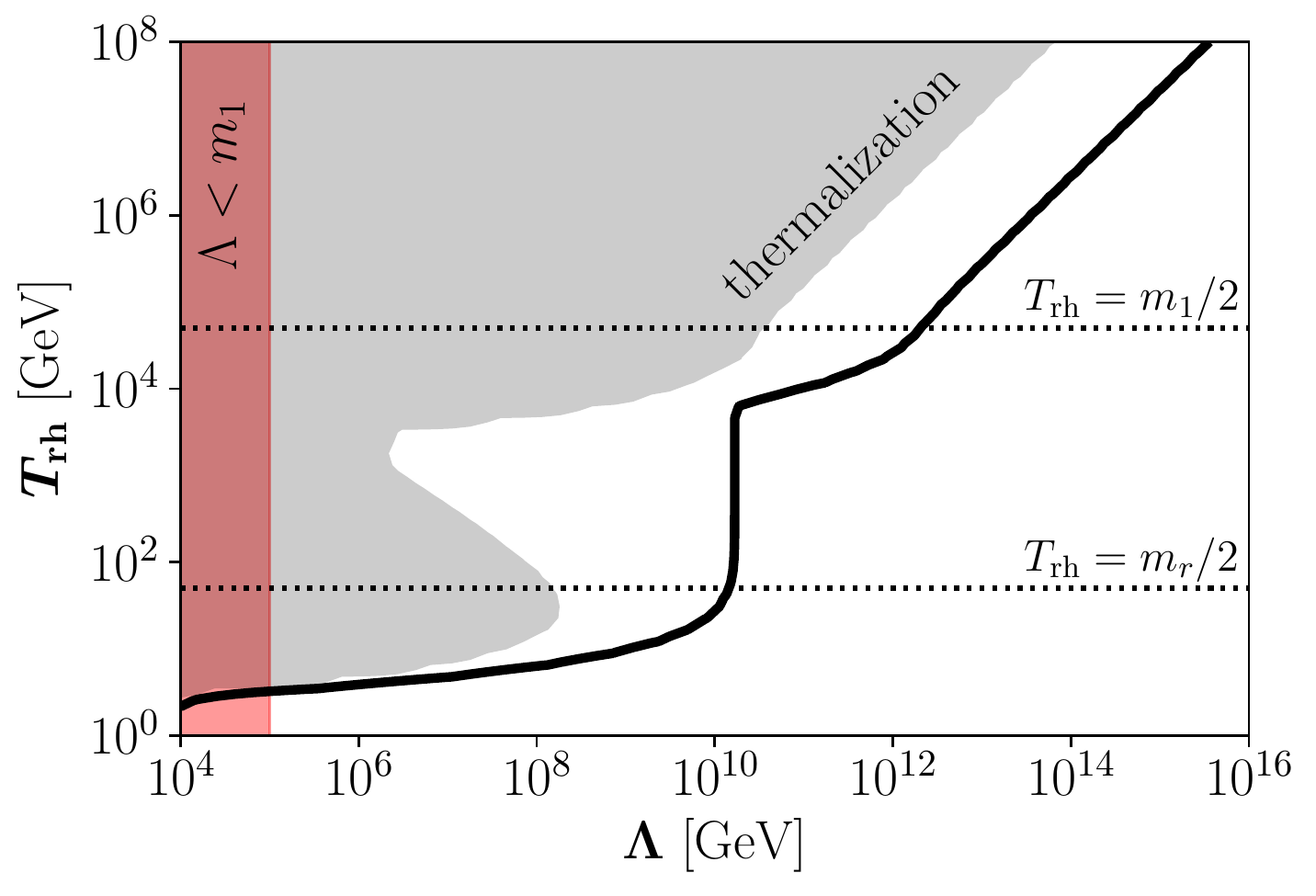}
	\caption{Direct freeze-in: Reheating temperature required to reproduce the experimentally observed DM abundance, $\Omega_\chi h^2$, 
	for $m_\chi=1$~MeV. Left panel: $\Trh$ as a function of $m_1$ for  $\Lambda=10^{11}$~GeV; right panel: $\Trh$ as a function of $\Lambda$ for $m_1=10^5$~GeV.
	In both panels, the radion mass is $m_r=m_1/10^3$.
	The gray-shaded areas are the regions where chemical equilibrium with the SM is reached (and freeze-in does not occur), whereas the red-shaded areas are the regions
	where $m_1>\Lambda$ and the EFT approach breaks down. Eventually, the two black-dotted lines give a visual understanding of the different regions in 
	eq.~\eqref{eq:yield_direct}. 
}
	\label{fig:directFI}
\end{center}
\end{figure}

Fig.~\ref{fig:directFI} shows the reheating temperature $\Trh$ required to reproduce the experimentally observed DM abundance, $\Omega_\chi h^2$, 
for a fixed value of the DM mass, $m_\chi=1$~MeV. 
In the left panel, we show $\Trh$ as a function of the first KK-graviton mass, $m_1$, for fixed $\Lambda = 10^{11}$~GeV; in the right panel, we show
$\Trh$ as a function of $\Lambda$ for fixed $m_1 = 10^5$ GeV. The radion mass has been chosen as $m_r=m_1/10^3$ (therefore, it is a variable parameter in the left panel, 
whereas it is a fixed one in the right panel).
In order to compute $\Trh$, the DM yield has been held fixed so that $m_\chi\,Y_0 = \Omega_\chi h^2 \, \frac{1}{s_0}\,\frac{\rho_c}{h^2} \simeq 4.3 \times 10^{-10}$~GeV, 
where $\rho_c \simeq 1.1 \times 10^{-5} \, h^2$~GeV/cm$^3$ is the critical energy density, $s_0\simeq 2.9\times 10^3$~cm$^{-3}$ is the entropy density at present 
and $\Omega_\chi h^2\simeq 0.12$~\cite{Aghanim:2018eyx}.
The gray-shaded areas are the regions where chemical equilibrium with the SM is reached and, therefore, where the freeze-in cannot occur and the analysis performed here 
 is not valid.
The black-dotted lines, representing $\Trh=m_1/2$ and $\Trh=m_r/2$, have been added for reference.
Eventually, the red-shaded areas ($m_1>\Lambda$) represent the regions for which the EFT approach breaks down.

For the sake of completeness, notice that the $s$-channel exchange of a (massless) graviton gives an irreducible contribution to the total DM relic abundance~\cite{Garny:2015sjg, Tang:2017hvq, Garny:2017kha, Bernal:2018qlk}.
However, due to the large hierarchy $\Lambda\ll M_P$, the contribution of the massless graviton is typically subdominant and can be disregarded.
The corresponding interaction rate density is given by:
\begin{equation}
	\gamma_\text{DM$\to$SM}\simeq 1.9\times 10^{-4}\frac{T^8}{M_P^4} \,,
\end{equation}
and, therefore, its contribution to the DM yield is:
\begin{equation}
	Y_0\simeq \frac{1.4\times 10^{-4}}{\gss}\sqrt{\frac{10}{\gs}} \left ( \frac{\Trh}{M_P} \right )^3 \, .
\end{equation}
We stress that this expression is a function of $\Trh$, only, being naturally independent from $\Lambda$ and the masses of the KK-gravitons and the radion. This contribution, indeed, comes
from 4-dimensional gravitons or, in the case considered here,  from the long distance (low-energy) limit of 5-dimensional gravitons (corresponding to the KK-graviton zero-mode). 
For example, for a DM mass $m_\chi=10$~TeV
it would only be relevant for reheating temperatures $\Trh \geq 10^{16}$ GeV, {\em i.e.} well above the range of $\Trh$ depicted in Fig.~\ref{fig:directFI}.

\subsection{Sequential Freeze-in}
In this case the DM abundance comes from decays of KK-gravitons or radions, previously produced via the freeze-in mechanism.
Such states are mainly generated by 2-to-2 annihilations or inverse decays (2-to-1) of SM particles. We will now review one by one the two possibilities.

\subsubsection{Via Annihilations}

KK-gravitons and radions with masses below the reheating temperature can be created {\it on-shell} in the early Universe via annihilations of two SM particles 
by the freeze-in mechanism. Once created, their decay products may contribute to the total DM relic abundance.
In fact, if the production cross-section is small enough to keep KK-gravitons and radions out of chemical equilibrium with the SM bath, and the evolution of the DM yield is largely dominated by their decays, eqs.~\eqref{eq:cosmo1} to~\eqref{eq:cosmo2} can be simplified to:
\begin{eqnarray}\label{eq:BEFIsecann}
	\frac{dY}{dT}&\simeq&\frac{\gamma_\text{KK$\to$SM}}{H\,\mathfrak{s}\,T}\left[\left(\frac{Y_K}{Y_K^\text{eq}}\right)^2-1\right]\,\text{BR}(\text{KK$\to$DM})
	              + \frac{\gamma_\text{r$ \to$SM}}{H\,\mathfrak{s}\,T}\left[\left(\frac{Y_r}{Y_r^\text{eq}}\right)^2-1\right] \, \text{BR}(\text{r$ \to$DM}) \nonumber\\
	&\simeq&-\frac{1}{H\,\mathfrak{s}\,T}\left[\gamma_\text{KK$\to$SM} \, \text{BR}(\text{KK$\to$DM})+\gamma_\text{r$ \to$SM} \, \text{BR}(\text{r$ \to$DM}) \right] \, ,
\end{eqnarray}
where the rates are:
\begin{eqnarray}
\label{eq:GammaK-SM}
	\gamma_\text{KK$\to$SM}(T)&\simeq& 4.8\times 10^4\frac{T^{16}}{\Lambda^4 m_n^8} \qquad \text{(for the $n^\text{th}$ KK-graviton)}\,, \\
	\gamma_\text{r$ \to$SM}(T)&\simeq& 2.2\times 10^{-4}\frac{T^{8}}{\Lambda^4} \, . \label{eq:Gammaannr-SM}
\end{eqnarray}
Notice that the $m_n^{-8}$ factor  in $\gamma_\text{KK $\to$ SM}$ (and, hence, the strong temperature dependence) comes from the polarization tensor of the KK-gravitons 
(as it was shown in Refs.~\cite{Lee:2013bua,Folgado:2019sgz} for spin-2 massive particles). Such a suppression is not present in the case of radions (that have spin 0).
The branching ratios into DM particles are:
\begin{eqnarray}
\label{eq:BRKK}
	\text{BR} (\text{KK$\to$DM}) & \simeq & \frac{z_n}{z_n+256}\,,\\
	\text{BR} (\text{r$ \to$DM})    & \simeq & \frac{z}{z+37}\,,
\end{eqnarray}
where
\begin{eqnarray}
	z_n &\equiv& \left (1-4\frac{m_\chi^2}{m_n^2}\right )^{5/2},\\
	z &\equiv& \sqrt{1 - 4 \frac{m_\chi^2}{m_r^2}} \left ( 1 + 2 \frac{m_\chi^2}{m_r^2} \right )^{2}.
\end{eqnarray}
The explicit expressions for annihilation rates and decay widths for KK-gravitons and the radion can be found in Appendix~\ref{app:annihil}.

Using a similar procedure to the one used in eq.~\eqref{eq:BEapproxDFI} and~\eqref{eq:BEFIsecann}, it is possible to find the following analytical solution:
\begin{equation}
\label{eq:yield_sfi_ann}
	Y_0 \simeq \frac{9.5\times 10^3}{\gss}\sqrt{\frac{10}{\gs}}\frac{M_P}{\Lambda^4 m_1^8} \left ( \frac{z_1}{z_1+256} \right ) 
	\,\Trh^{11} + \frac{1.6\times 10^{-4}}{\gss}\sqrt{\frac{10}{\gs}}\frac{M_P}{\Lambda^4 } \left ( \frac{z}{z+37} \right ) \,\Trh^{3}  \, .
\end{equation}
Notice that in eq.~(\ref{eq:yield_sfi_ann}) only the lightest KK-graviton is taken into account. This is a consequence of the strong suppression with the KK-graviton mass $m_n$
in eq.~(\ref{eq:GammaK-SM}). Even if all of the KK-gravitons do contribute to the total DM density, the only relevant contribution is given by the lightest state.
For the previous analysis to be valid, the KK-gravitons and the radion must be out of chemical equilibrium with the SM bath, 
which corresponds to the conditions $\gamma_\text{KK$\to$SM}\ll n_K^\text{eq}\,H$ and $\gamma_\text{r$\to$SM}\ll n_r^\text{eq}\,H$.
The reheating temperature in this limit satisfies the tightest of the following conditions (depending on the mass of the lightest KK-graviton, $m_1$):
\begin{equation}
	\Trh \ll  \min \left ( 0.3\left[\sqrt{\frac{\gs}{10}}\frac{\Lambda^4\,m_1^8}{M_P}\right]^{1/11}; \, 8.3\left[\sqrt{\frac{\gs}{10}}\frac{\Lambda^4}{M_P}\right]^{1/3} \right).
\end{equation}

\begin{figure}[t]
\begin{center}
\includegraphics[height=0.33\textwidth]{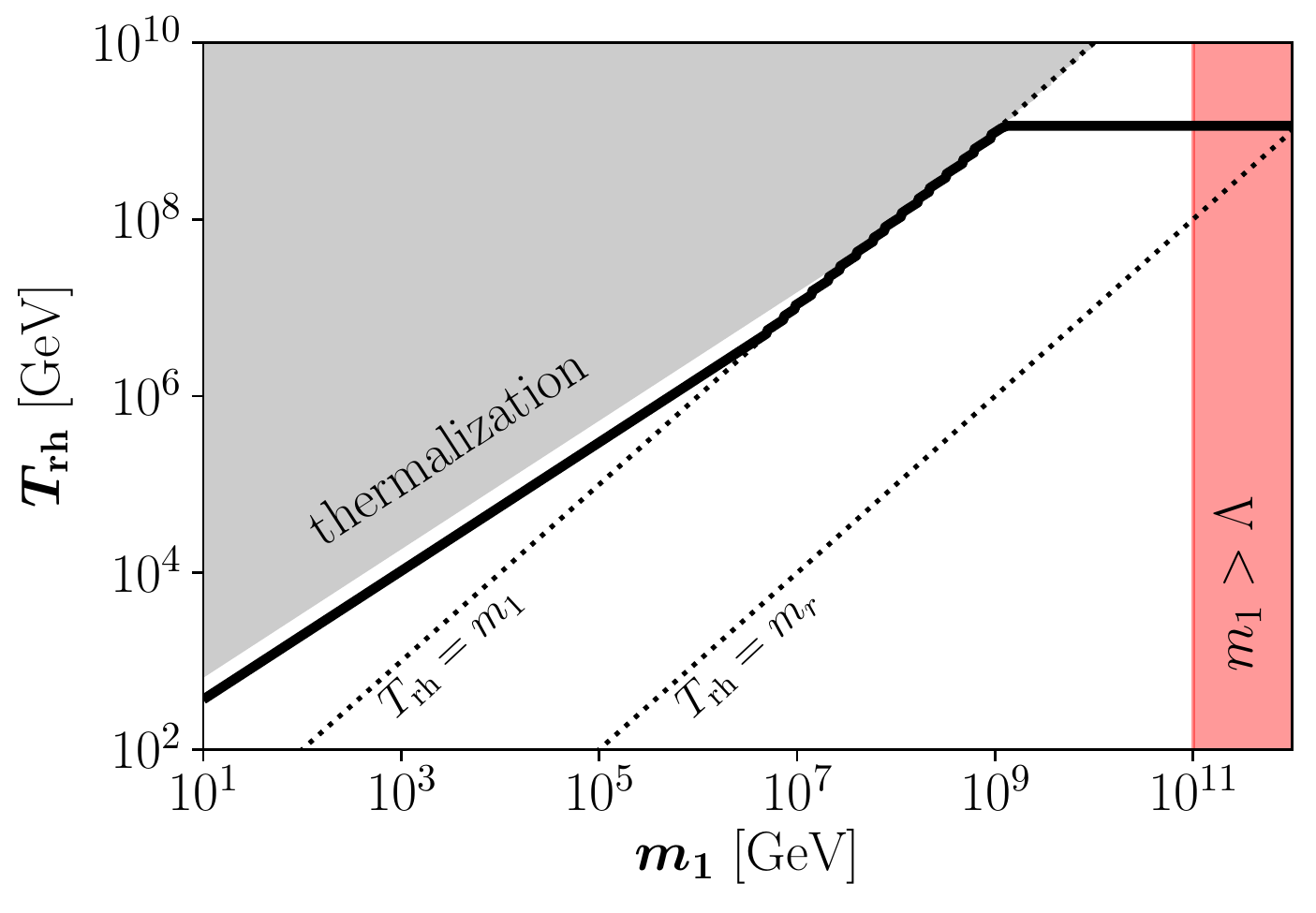}
\includegraphics[height=0.33\textwidth]{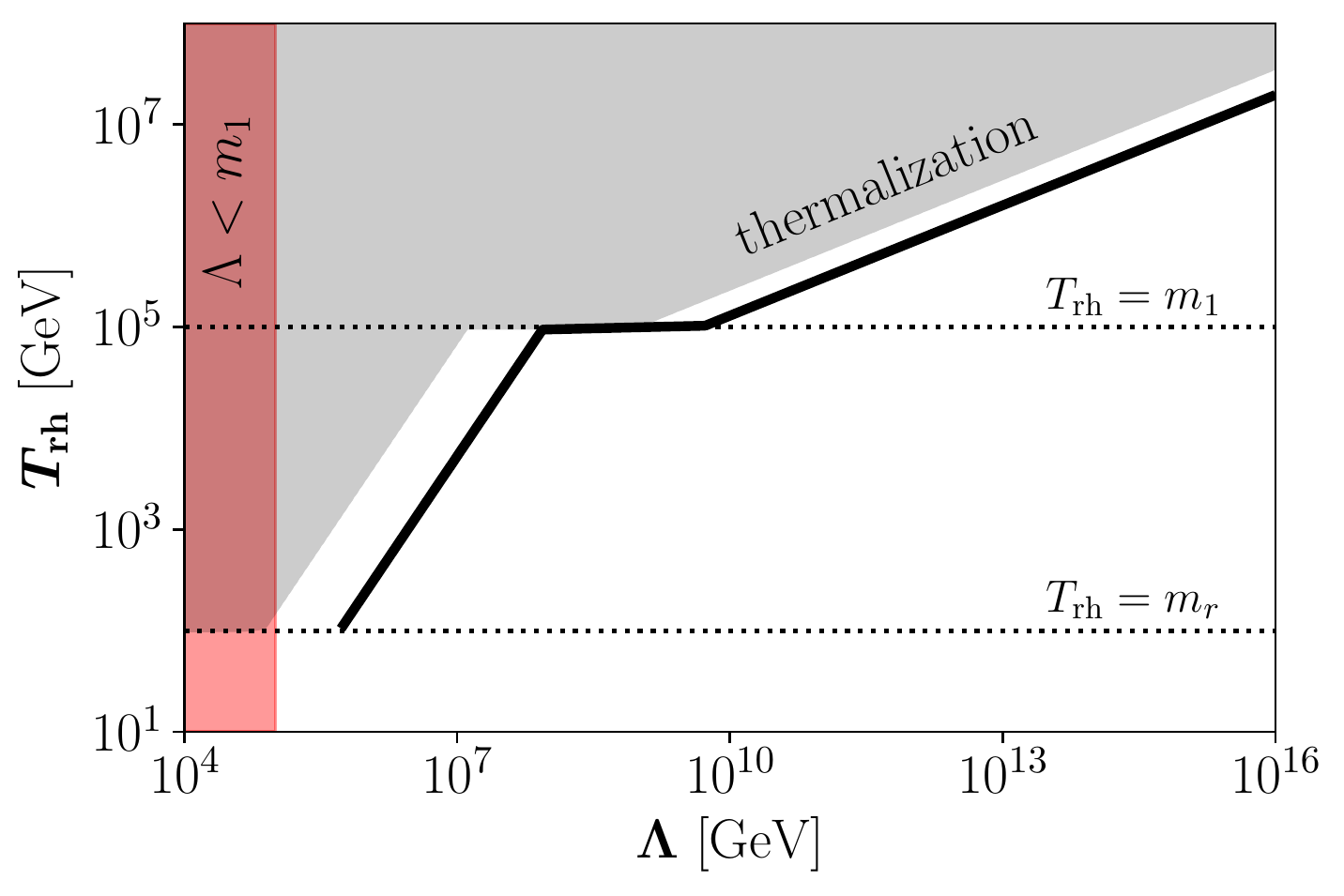}
	\caption{Sequential freeze-in via annihilations:
	Reheating temperature required to reproduce the experimentally observed DM abundance, $\Omega_\chi h^2$, 
	for $m_\chi=1$~MeV. Left panel: $\Trh$ as a function of $m_1$ for  $\Lambda=10^{11}$~GeV; right panel: $\Trh$ as a function of $\Lambda$ for $m_1=10^5$~GeV.
	In both panels, the radion mass is $m_r=m_1/10^3$.
	The gray-shaded areas are the regions where chemical equilibrium with the SM is reached (and freeze-in does not occur), whereas the red-shaded areas are the regions
	where $m_1>\Lambda$ and the EFT approach breaks down. Eventually, the two black-dotted lines give a visual understanding of the different regions in 
	eq.~\eqref{eq:yield_sfi_ann}. 
}
	\label{fig:sequentialFIann}
\end{center}
\end{figure}

Fig.~\ref{fig:sequentialFIann} shows the reheating temperature $\Trh$ required to reproduce the observed DM abundance 
for a fixed value of the DM mass, $m_\chi=1$~MeV.
As in Fig.~\ref{fig:directFI}, in the left panel we show $\Trh$ as a function of the first KK-graviton mass, $m_1$, for fixed $\Lambda = 10^{11}$ GeV; in the right panel, we show
$\Trh$ as a function of $\Lambda$ for fixed $m_1 = 10^5$ GeV. The relation between the radion mass $m_r$ and the lightest KK-graviton mass, $m_1$ is, again,  $m_r=m_1/10^3$.
The black-dotted lines indicate $\Trh=m_1$ and $\Trh=m_r$. Eventually, the gray- and red-shaded areas are the regions where chemical equilibrium with the SM is reached, 
and where the EFT approach breaks down (as $m_1>\Lambda$), respectively.

For $\Trh < m_r$, on-shell KK gravitons and radions are not produced in the early Universe, and therefore this mechanism can not account for the DM relic abundance.
If $m_r < \Trh < m_1$, only radions are created. In this region $\Trh$ is independent on $m_1$ (and therefore on $m_r$) due to the fact that the interaction rate in 
eq.~\eqref{eq:Gammaannr-SM} does not depend on $m_r$, as it can be seen in the left panel of Fig.~\ref{fig:sequentialFIann}.
Now, if $\Trh > m_1$ the KK-gravitons are also produced and their decay dominate the DM production. The reheating temperature needed to reproduce the observed value 
of $\Omega_\chi h^2$ in this region is very near to the border of the gray-shaded area for which the DM is in equilibrium with SM particles and freeze-in does not occurs
(remember, though, the log-log scale of the plots).

\subsubsection{Via Inverse Decays}
Alternatively , frozen-in KK-gravitons and radions are also created {\it on-shell} via inverse decays of SM particles (a 2-to-1 process), and subsequently they can decay into DM particles.
Within the same approximations as in the previous subsection, i.e.,  assuming that KK-gravitons and radions are produced out of chemical equilibrium from the SM bath via inverse-decays, and the evolution of the DM yield is largely dominated by their decays, eqs.~\eqref{eq:cosmo1} to~\eqref{eq:cosmo2} can be simplified to:
\begin{eqnarray}\label{eq:BEFIsecdec}
	\frac{dY}{dT}&\simeq&\frac{\gamma^d_\text{KK$\to$SM}}{H\,\mathfrak{s}\,T}\left[\frac{Y_K}{Y_K^\text{eq}}-1\right]\,\text{BR}(\text{KK$\to$DM})
	              + \frac{\gamma^d_\text{r$ \to$SM}}{H\,\mathfrak{s}\,T}\left[\frac{Y_r}{Y_r^\text{eq}}-1\right]\,\text{BR}(\text{r$ \to$DM}) \nonumber\\
	&\simeq&-\frac{1}{H\,\mathfrak{s}\,T}\left[\gamma^d_\text{KK$\to$SM}\,\text{BR}(\text{KK$\to$DM}) + \gamma^d_\text{r$ \to$SM}\,\text{BR}(\text{r$ \to$DM}) \right],
\end{eqnarray}
where the interaction rate densities for decays are defined by:
\begin{equation}
        \gamma^d(T)=\frac{m^2\,T}{2\pi^2}K_1\left(\frac{m}{T}\right)\,\Gamma\,,
\end{equation}
with $\Gamma$ the decay width obtained by summing (rather than averaging) over the degrees of freedom of the decaying particle.
Using eqs.~\eqref{eq:decKKtoSM} and~\eqref{eq:decrtoSM} we get, then:
\begin{eqnarray}
	\gamma^d_\text{KK$\to$SM}&\simeq&\frac{73}{480\pi^3} \frac{m_n^5\,T}{\Lambda^2}\,K_1\left(\frac{m_n}{T}\right),\\
	\gamma^d_\text{r$ \to$SM}&\simeq&\frac{37}{384\pi^3} \frac{m_r^5\,T}{\Lambda^2}\,K_1\left(\frac{m_r}{T}\right).
\end{eqnarray}

Eq.~\eqref{eq:BEFIsecdec} admits the following approximate analytical solution:
\begin{equation}
\label{eq:YieldSecInv1}
	Y_0\simeq\sum_n\frac{5.6\times 10^{-2}}{\gss}\sqrt{\frac{10}{\gs}}\frac{M_P\,m_n}{\Lambda^2} \left ( \frac{z_n}{z_n+256} \right ) 
	+ \frac{3.5\times 10^{-2}}{\gss}\sqrt{\frac{10}{\gs}}\frac{M_P\,m_r}{\Lambda^2} \left ( \frac{z}{z+37} \right ) \, .
\end{equation}
In this case, most of the DM production happens at $T\simeq m_n/2.5$ and $T\simeq m_r/2.5$ for KK-gravitons and radions, respectively. However, 
the sum over KK-modes should be performed up to KK-graviton states with mass below the reheating temperature, $m_n<\Trh$. 
For this reason, the total contribution due to the decay of KK-gravitons explicitly depends on $\Trh$ (whereas the second term in eq.~(\ref{eq:YieldSecInv1}) does not depend on it):
\begin{equation}\label{eq:YieldSecInv}
	Y_0\simeq\frac{2.2\times 10^{-4}}{\gss}\sqrt{\frac{10}{\gs}}\frac{M_P\,\Trh^2}{m_1\,\Lambda^2} + \frac{3.5\times 10^{-2}}{\gss}\sqrt{\frac{10}{\gs}}\frac{M_P\,m_r}{\Lambda^2}
	\left ( \frac{z}{z+37} \right ) \,.
\end{equation}

Again, for the KK-gravitons and the radions to be out of chemical equilibrium with the SM bath one needs to guarantee that $\gamma^d_\text{KK$\to$SM}\ll n_K^\text{eq}\,H$ 
and $\gamma^d_\text{r$\to$SM}\ll n_r^\text{eq}\,H$. The reheating temperature in this limit satisfies the tightest of the following conditions (depending on the mass of the lightest KK-graviton, $m_1$):
\begin{equation}
	\Trh \ll \min \left ( 0.34 \left [ \sqrt{\frac{10}{\gs}} \, \frac{M_P\,m_1^4}{\Lambda^2}\right]^{1/3}     ;\, 
	                            0.29 \left [ \sqrt{\frac{10}{\gs}} \, \frac{M_P\,m_r^4}{\Lambda^2}\right]^{1/3} \right )       .
\end{equation}

\begin{figure}[t]
\begin{center}
\includegraphics[height=0.33\textwidth]{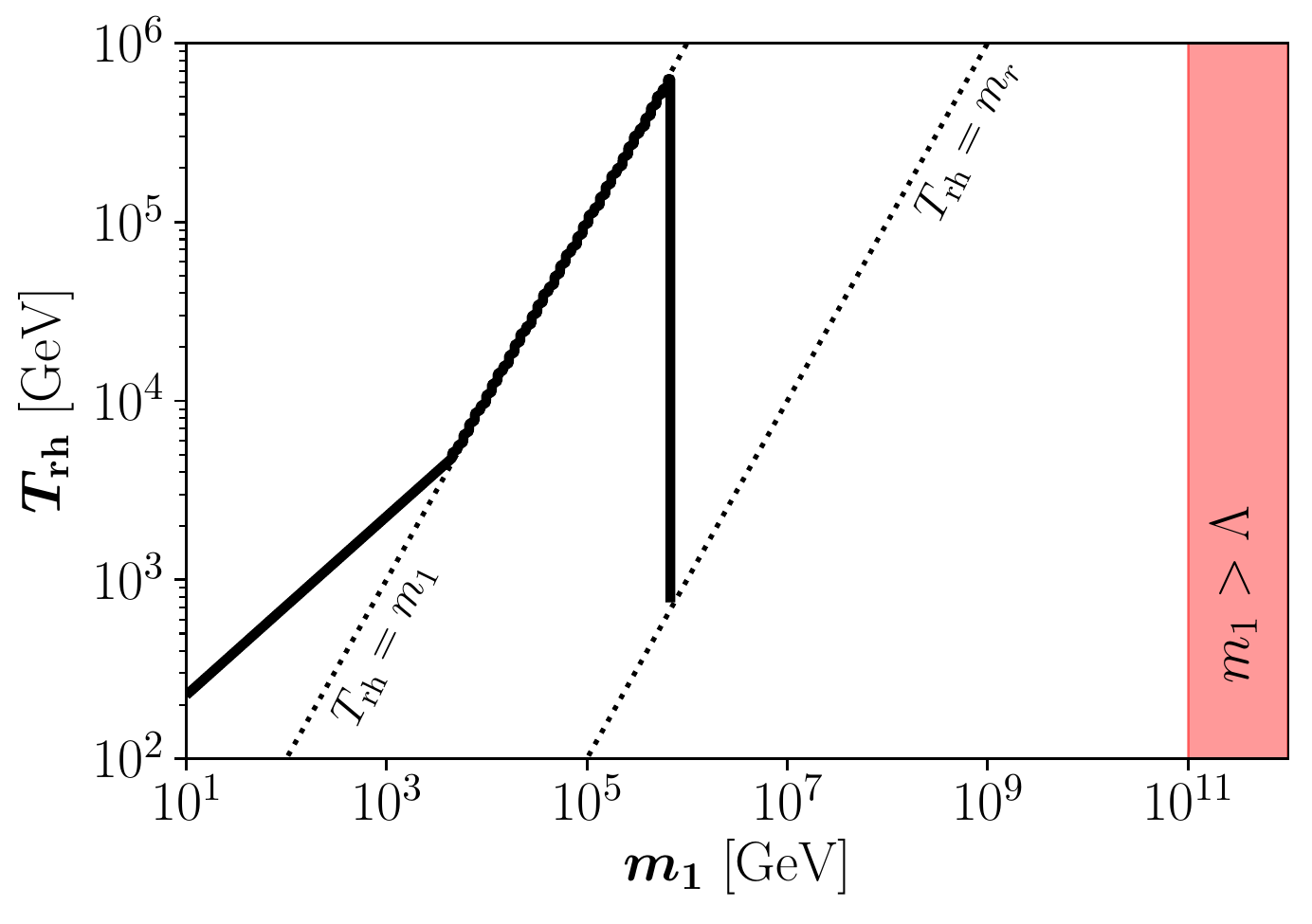}
\includegraphics[height=0.33\textwidth]{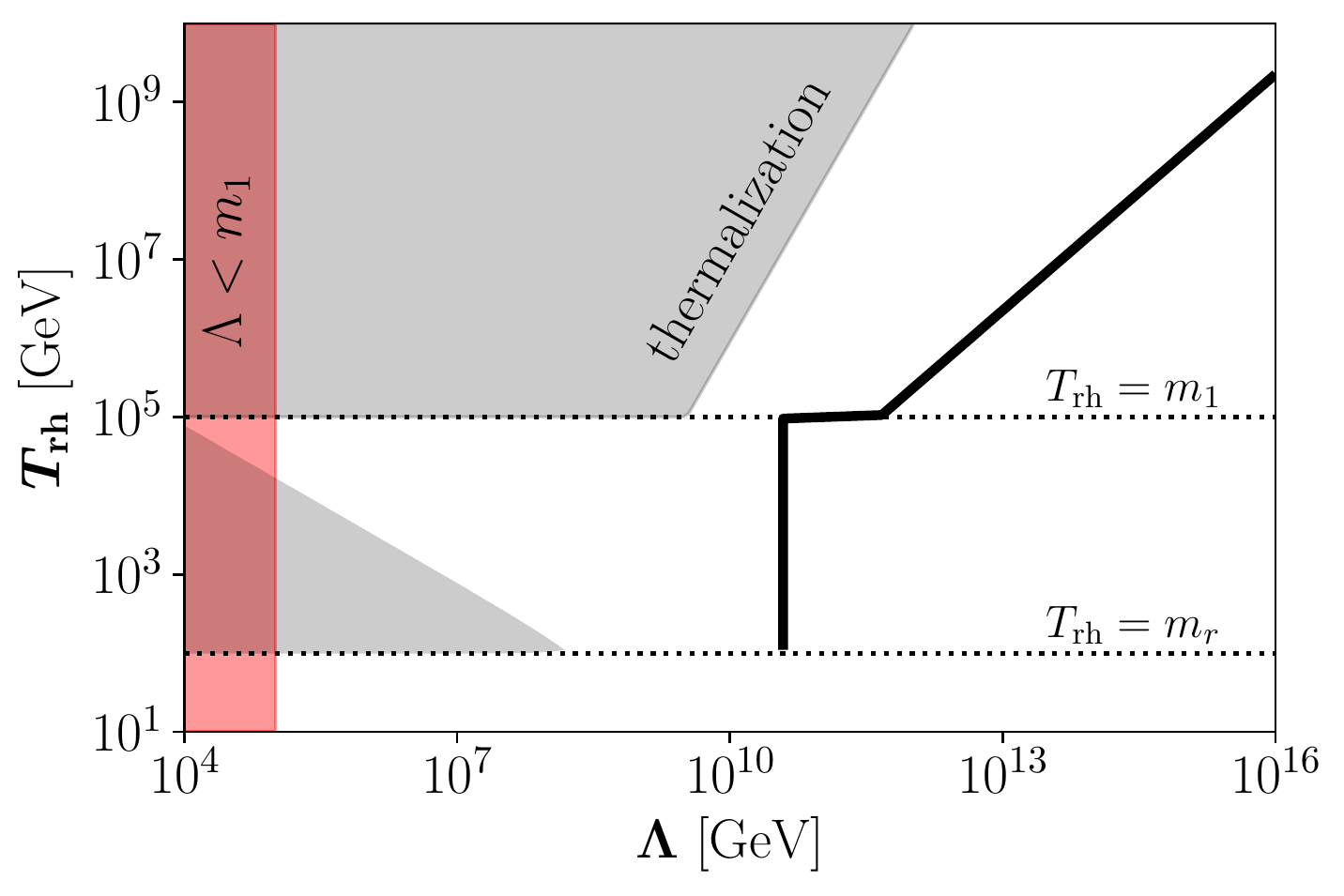}
	\caption{Sequential freeze-in via inverse decays: 
	Reheating temperature required to reproduce the experimentally observed DM abundance, $\Omega_\chi h^2$, 
	for $m_\chi=1$~MeV. Left panel: $\Trh$ as a function of $m_1$ for  $\Lambda=10^{11}$~GeV; right panel: $\Trh$ as a function of $\Lambda$ for $m_1=10^5$~GeV.
	In both panels, the radion mass is $m_r=m_1/10^3$.
	The gray-shaded areas are the regions where chemical equilibrium with the SM is reached (and freeze-in does not occur), whereas the red-shaded areas are the regions
	where $m_1>\Lambda$ and the EFT approach breaks down. Eventually, the two black-dotted lines give a visual understanding of the different regions in 
	eq.~\eqref{eq:yield_direct}.
	}
	\label{fig:sequentialFIdec}
\end{center}
\end{figure}
Fig.~\ref{fig:sequentialFIdec} shows the reheating temperature $\Trh$ required to reproduce the observed DM abundance, $\Omega_\chi h^2$, 
for a fixed value of the DM mass, $m_\chi=1$~MeV.
Again, in the left panel we show $\Trh$ as a function of the first KK-graviton mass, $m_1$, for fixed $\Lambda = 10^{11}$ GeV; in the right panel, we show
$\Trh$ as a function of $\Lambda$ for fixed $m_1 = 10^5$ GeV. The radion mass has been chosen as $m_r=m_1/10^3$. 
The black-dotted lines indicate $\Trh=m_1$ and $\Trh=m_r$. 
The gray- and red-shaded areas are the regions where chemical equilibrium with the SM is reached,\footnote{Notice that in the left panel the gray-shaded area is absent as the region 
for which the DM is in equilibrium with SM particles is outside of the considered range.} 
and where the EFT approach breaks down (as $m_1>\Lambda$), respectively. 

As in the case of sequential freeze-in via annihilation, for $\Trh < m_r$ on-shell KK-gravitons and radions are not produced in the early Universe and, therefore, 
this mechanism can not account for the DM relic abundance below the $\Trh = m_r$ black-dotted line. In the region $m_r < \Trh < m_1$, only radions are created and, 
in this case, the DM yield is independent on $\Trh$ (as the second term in eq.~\eqref{eq:YieldSecInv} does not depend on $\Trh$). 
This can be clearly seen in Fig.~\ref{fig:sequentialFIdec}. For $\Trh > m_1$,  the KK-graviton states are also produced. Their decay eventually dominate the DM production and
the reheating temperature is proportional to $\sqrt{m_1}$ (left panel) or $\Lambda$ (right panel).

\begin{figure}[t]
\begin{center}
\includegraphics[height=0.33\textwidth]{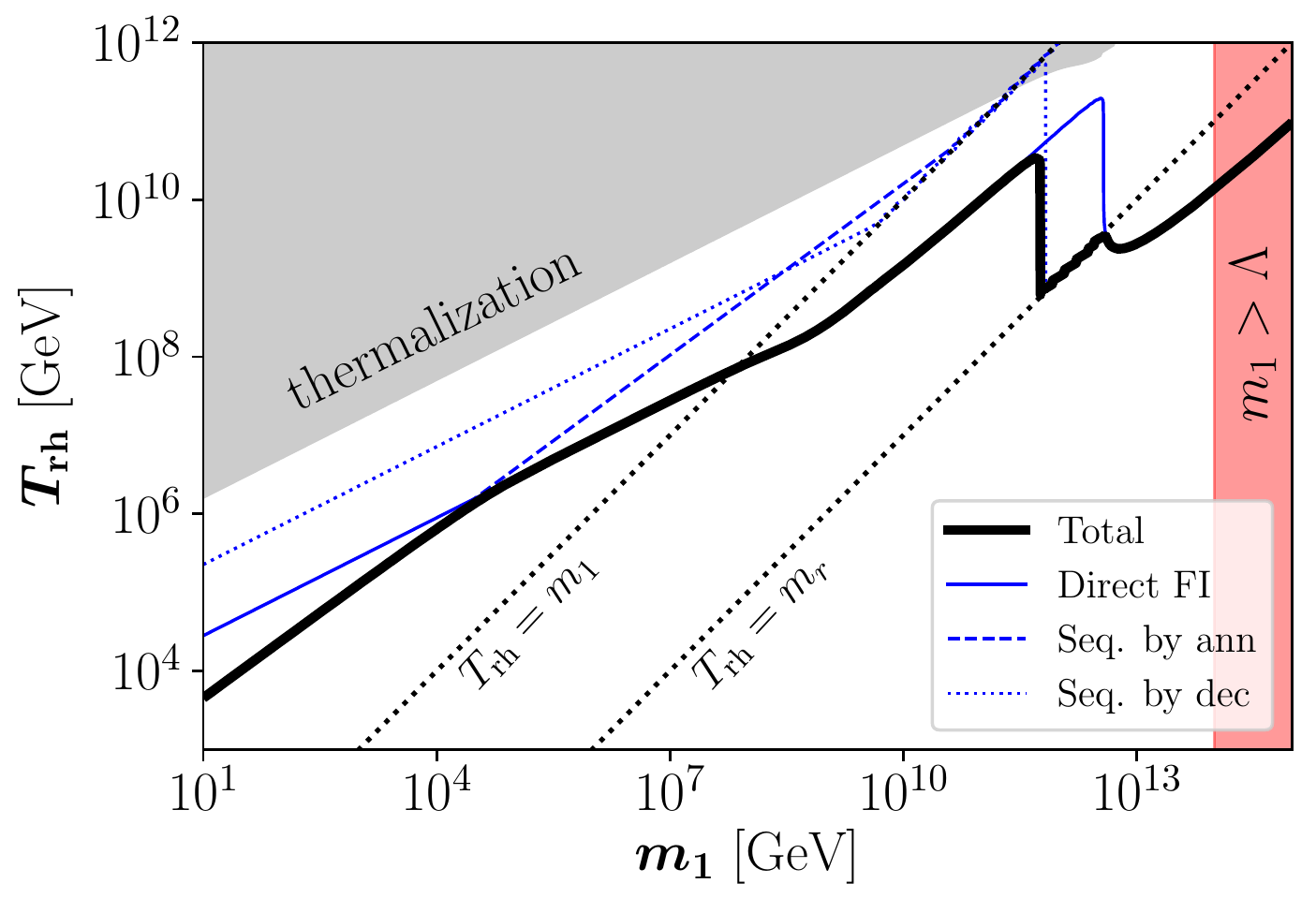}
\includegraphics[height=0.33\textwidth]{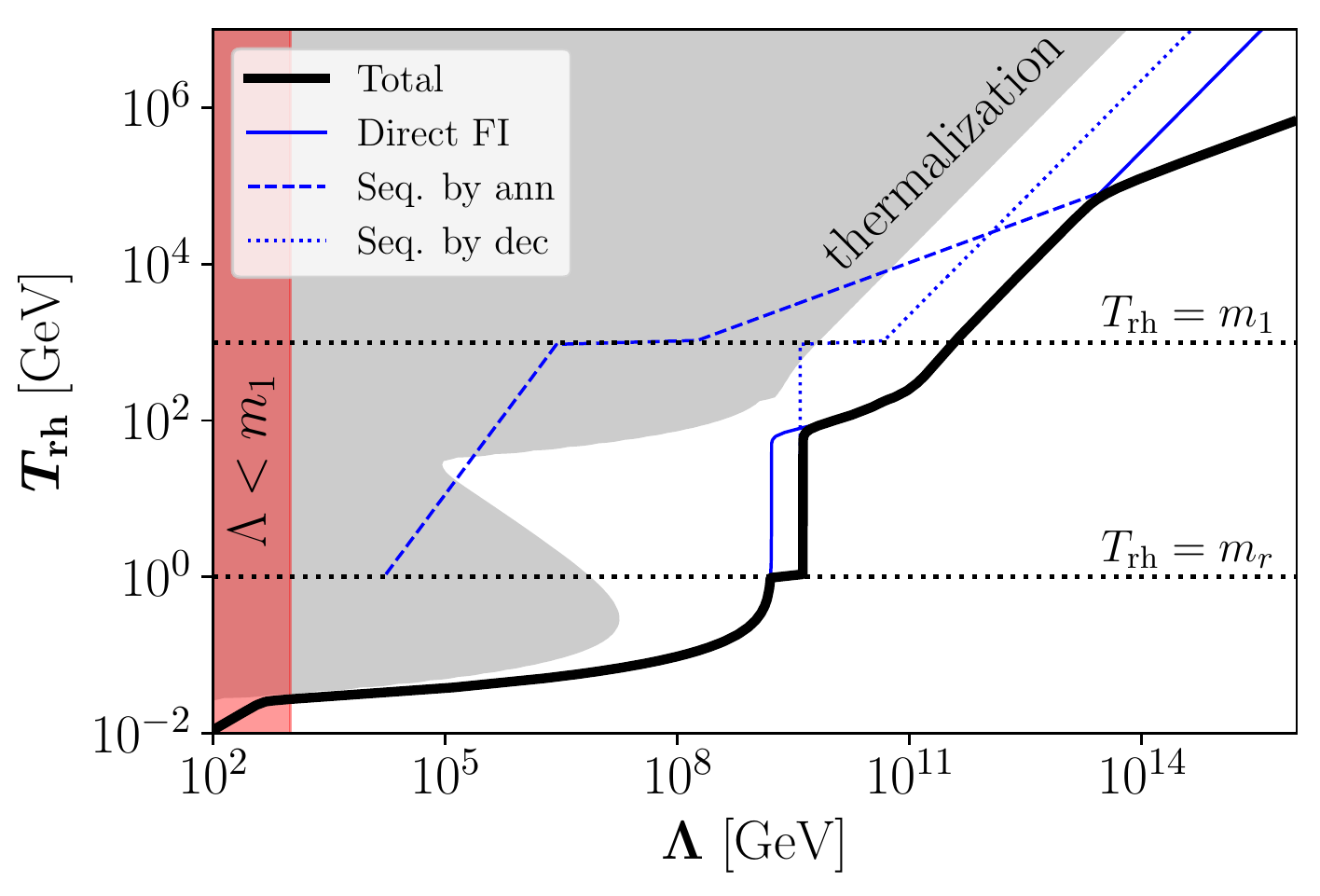}
\includegraphics[height=0.33\textwidth]{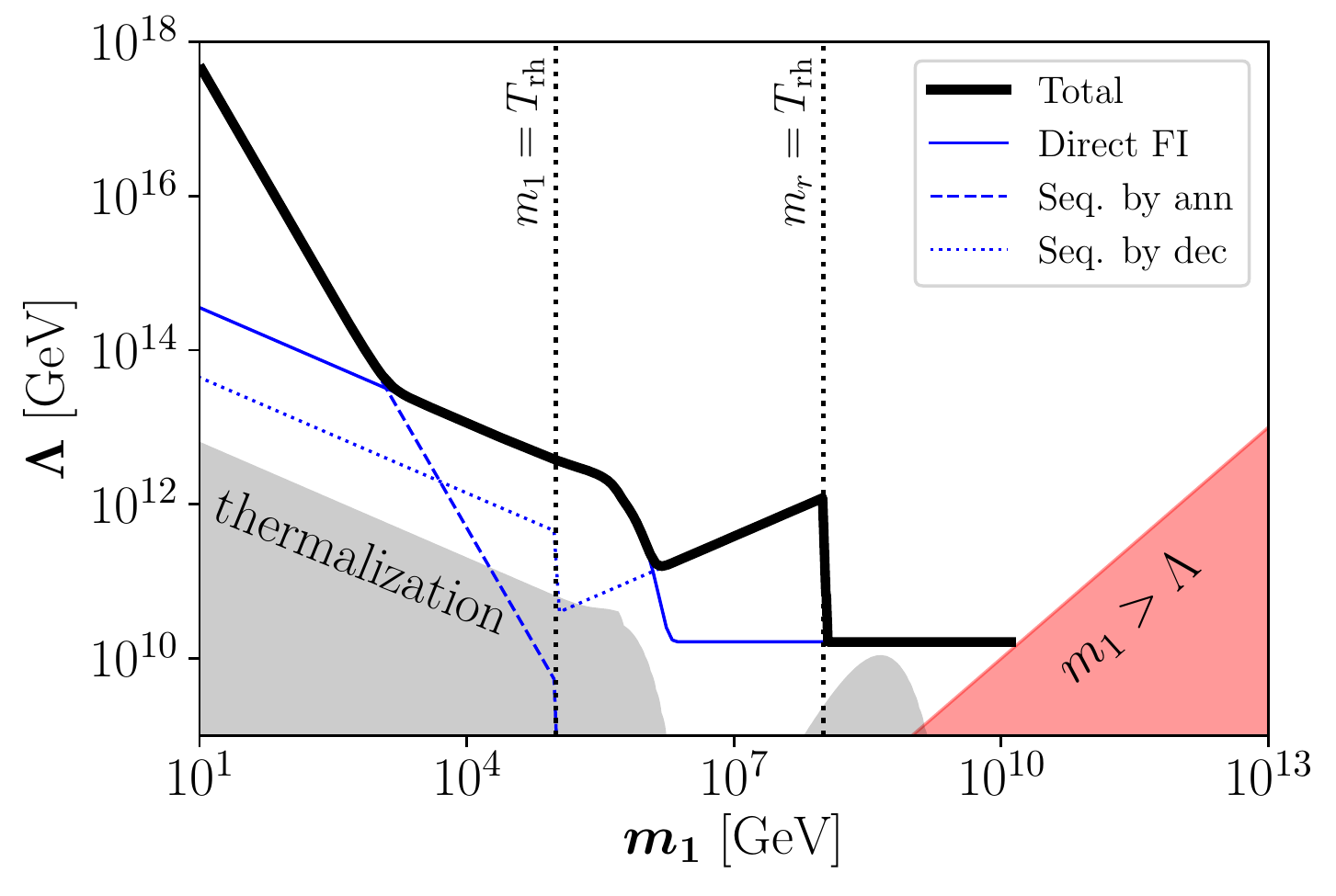}
	\caption{
	Reheating temperature required to reproduce the experimentally observed DM abundance, $\Omega_\chi h^2$, 
	for $m_\chi=1$~MeV, taking into account all possible DM production mechanisms (thick black lines).
	Upper left panel: $\Trh$ as a function of $m_1$ for  $\Lambda=10^{14}$~GeV; Upper right panel: $\Trh$ as a function of $\Lambda$ for $m_1=10^3$~GeV;
	Lower panel: correlation between $\Lambda$ and $m_1$ for $\Trh=10^5$~GeV. In all panels, $m_r=m_1/10^3$, whereas as always
        the gray- and red-shaded areas are the regions where chemical equilibrium with the SM is reached and where  the EFT approach breaks down, respectively.
        The two black-dotted lines represent the conditions $\Trh = m_1$ and $\Trh = m_r$. The light blue solid, dashed and dotted lines represent the contributions
        from direct freeze-in, sequential freeze-in via annihilation and sequential freeze-in via inverse decay, respectively (as explained in the legend). 
}
	\label{fig:all}
\end{center}
\end{figure}
So far, each individual production channel has been studied separately. Fig.~\ref{fig:all} depicts the parameter space favored by the observed DM abundance for $m_\chi=1$~MeV and 
$\Lambda=10^{14}$~GeV as a function of $m_1$ (upper left panel), or $m_1=10^3$~GeV as a function of $\Lambda$ (upper right panel), 
taking into account {\it all} of the three DM production mechanisms described previously ({\em i.e.} direct production, sequential production via annihilation and sequential production via inverse decay).
The thin blue lines correspond to the partial contributions of each of the mechanisms, whereas the black thick line to the total abundance.
Eventually, in the lower panel we show the correlation between $\Lambda$ and $m_1$ at a fixed value of the reheating temperature required to achieve the observed DM abundance, $\Trh=10^5$~GeV.
In all panels, the radion mass is related to the first KK-graviton mass as $m_r=m_1/10^3$. As always, the gray- and red-shaded areas represent the regions where chemical equilibrium 
between DM and the SM particles is reached and where the EFT breaks down  since $m_1>\Lambda$, respectively.

\begin{figure}[t]
\begin{center}
\includegraphics[height=0.5\textwidth]{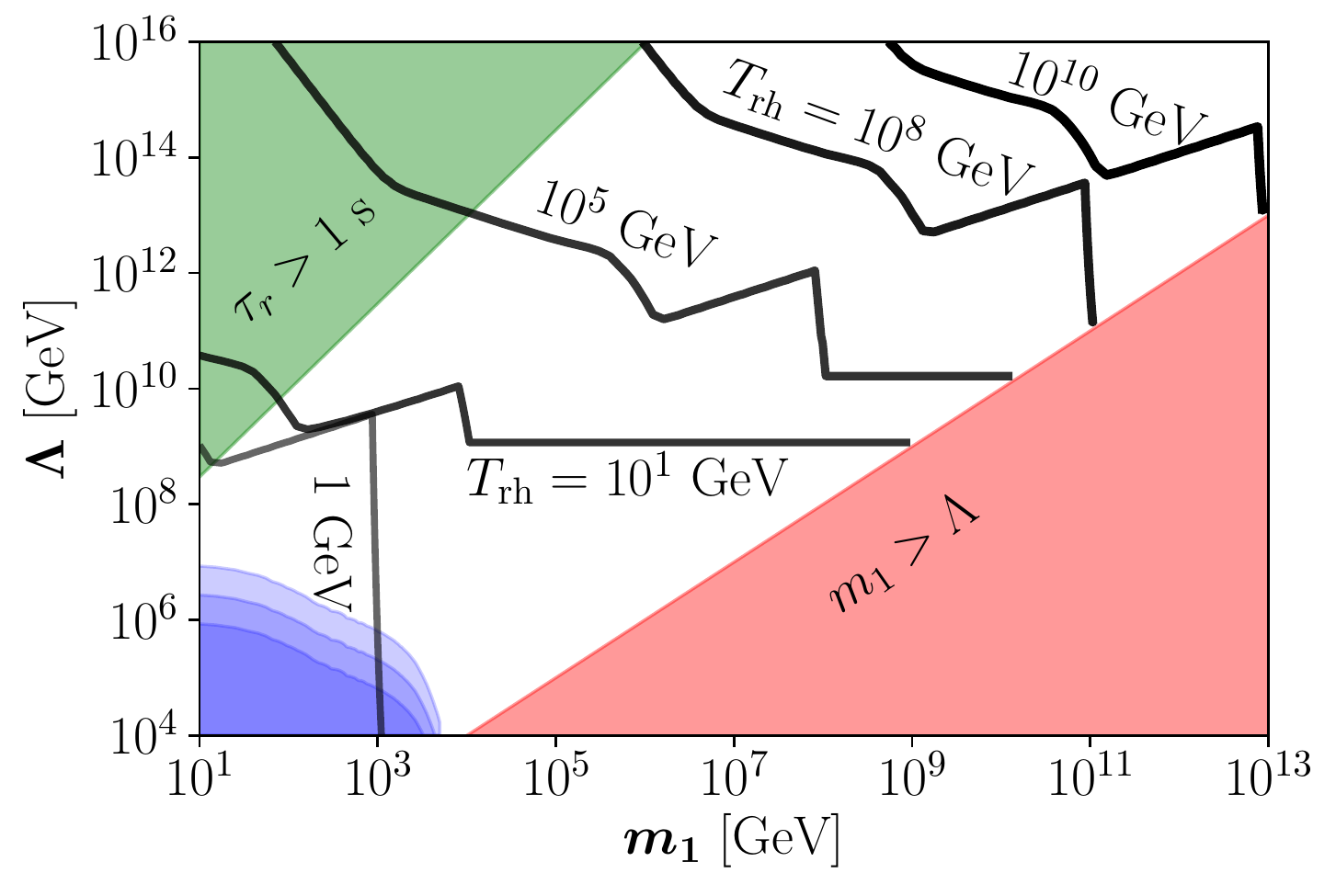}
	\caption{Parameter space required to reproduce the observed DM abundance for $m_\chi=1$~MeV and $m_r=m_1/10^3$, for several values of $\Trh$.
	The blue areas are excluded by resonant searches at LHC and represent the current bound and our prospects for the LHC Run-3 and the High-Luminosity LHC in the $\gamma \, \gamma$ channel~\cite{Aaboud:2017buh,Aaboud:2017yyg}, see text.
	The green corner corresponds to radion lifetimes longer than 1~s.
	In the red area ($m_1>\Lambda$) the EFT approach breaks down.
}
	\label{fig:all2}
\end{center}
\end{figure}
Finally, in Fig.~\ref{fig:all2} we show the correlation between $\Lambda$ and $m_1$ required to reproduce the observed DM abundance for $m_\chi=1$~MeV and $m_r=m_1/10^3$
for several representative values of the reheating temperature, $\Trh=1$, $10$, $10^5$, $10^8$ and $10^{10}$~GeV (notice that the range of $\Lambda$ plotted in Fig.~\ref{fig:all2} 
differs from that in the lower panel of Fig.~\ref{fig:all}).
Let us note that the lines corresponding to $\Trh=1$ and 10~GeV overlap when $m_1\simeq 10^3$~GeV and $\Lambda\simeq 10^9$~GeV.
This can be understood seeing that in that region, the DM relic abundance is mainly generated by sequential freeze-in via inverse decays of the radion and is therefore independent of $\Trh$, see eq.~\eqref{eq:YieldSecInv1}.
The red-shaded area, as always, represents the region where the EFT approach breaks down. On the other hand, 
the upper left green corner corresponds to radion lifetimes higher than 1~s, potentially problematic for BBN (all the KK-graviton states are heavier than the radion and therefore will naturally have shorter lifetimes). Eventually, the blue-shaded regions depict present and future experimental bound coming from resonance searches at the LHC.
The proton-proton collision can generate resonant KK-gravitons that later decay into SM particles. ATLAS and CMS put bounds over these processes in $\gamma \gamma$ and lepton-lepton channels 
as a function of the mass of the resonance (the lightest KK-graviton). These bounds can be translated into limits over $\Lambda$ as a function of the mass of the first graviton $m_1$.
The present bounds (dark blue) come from the resonant searches at LHC with 36~fb$^{-1}$~\cite{Aaboud:2017buh} and~\cite{Aaboud:2017yyg}, whereas future bounds 
are estimated assuming 300~fb$^{-1}$ (medium blue) and 3000~fb$^{-1}$ (light blue) for the LHC Run-III and High-Luminosity LHC, respectively.
Notice that in this plot we do not show the gray-shaded region for which DM is in equilibrium with SM particles (where freeze-in does not occur), as we should draw a different region for each
value of $\Trh$.

\subsection{Beyond the Sudden Decay Approximation of the Inflaton}

While reheating is commonly approximated as an instantaneous event, the decay of the inflaton into SM radiation is a continuous process~\cite{Scherrer:1984fd}.
Away from this approximation for reheating, the bath temperature may rise to a value $\Tmax$ which exceeds $\Trh$~\cite{Giudice:2000ex}.
It is plausible that the DM relic density may be established during this reheating period, in which case its abundance will significantly differ from freeze-in calculations assuming radiation domination.
In particular, it has been observed that if the DM is produced during the transition from matter to radiation domination via an interaction rate that scales like $\gamma(T)\propto T^n$, for $n>12$ the DM abundance is enhanced by a boost factor proportional to $(\Tmax/\Trh)^{n-12}$~\cite{Garcia:2017tuj}, whereas for $n\leq 12$ the difference between the standard UV freeze-in calculation differ 
only by an $\mathcal{O}(1)$ factor from calculations taking into account non-instantaneous reheating.
More recently, it has been highlighted that the critical mass dimension of the operator at which the instantaneous decay approximation breaks down depend on the equation of state $\omega$, or equivalently, to the shape of the inflationary potential at the reheating epoch~\cite{Bernal:2019mhf, Garcia:2020eof, Bernal:2020qyu}.
Therefore, the exponent of the boost factor becomes $(\Tmax/\Trh)^{n-n_c}$ with $n_c \equiv 6+2\, \left ( \frac{3-\omega}{1+\omega} \right ) $, 
showing a strong dependence on the equation of state~\cite{Bernal:2019mhf}.
Subsequent papers have explored the impact of this boost factor in specific models~\cite{Chen:2017kvz, Bernal:2018qlk, Bhattacharyya:2018evo, Chowdhury:2018tzw, Kaneta:2019zgw, Banerjee:2019asa, Chanda:2019xyl, Baules:2019zwk, Dutra:2019xet, Dutra:2019nhh, Mahanta:2019sfo, Cosme:2020mck}.
Finally, another way for enhancing the DM abundance occurs in cosmologies where inflation is followed by an epoch dominated by a fluid stiffer than radiation.
In such scenarios, even a small radiation abundance, produced for instance by instantaneous preheating effects, will eventually dominate the total energy density of the Universe without the need for a complete inflaton decay.
In particular, a strong enhancement if DM production happens via interaction rates with temperature dependence higher that $n_c=6$~\cite{Bernal:2020bfj}.

The present model of KK FIMP DM in warped extra-dimensions features processes where the interaction rate has a particularly strong temperature dependence, the most relevant ones being: $i)$ the DM annihilation into SM states for reheating temperatures much lower than the radion mass $\gamma_\text{DM $\to$ SM}(T)\propto T^{12}$; $ii)$ the same process near the resonances 
$\Trh\simeq m_r/2$ and $\Trh\simeq m_1/2$, where $\gamma_\text{DM $\to$ SM}(T)\propto T\,K_1(m_i/T)$ (with $m_i$ being the radion or the lightest KK-graviton mass, respectively);  
and $iii)$ the KK-graviton annihilation into SM particles $\gamma_\text{KK $\to$ SM}(T)\propto T^{16}$.
In these regimes, the non-instantaneous decay of the inflaton is expected to generate a strong boost factor to the DM yield, which translates into a reduction of the reheating temperature 
required to match the observed DM relic abundance.
As the precise determination of such boost factors depends on the details of the inflationary model (in particular on the energy density carried by the inflaton and its equation 
of state parameter previous to its decay), it is beyond the scope of this study.


\section{Conclusions}
\label{sec:conclusions}

Dark Matter (DM) is typically assumed to be made of weakly interacting massive particles (WIMPs), produced in the early Universe via the freeze-out mechanism.
Freeze-out occurs if the interactions between DM and SM particles are strong enough to bring them into chemical equilibrium.
However, if these interaction rates were never strong enough, the observed DM relic abundance could still have been produced by non-thermal processes, like the freeze-in mechanism.
In that case, DM is called a feebly interacting massive particle (FIMP).

In a warped extra-dimensional scenario, DM could naturally be a FIMP, if the effective gravitational scale $\Lambda$ is much higher than the electroweak scale.
In this case, DM is produced in two main ways:
$i)$ promptly by annihilations of SM states via the $s$-channel exchange of KK-gravitons and radions, i.e. the so called {\it direct freeze-in}, and
$ii)$ by decays of KK-gravitons or radions, previously produced by annihilations or inverse decays of SM particles via direct freeze-in. This scenario has been doubted {\it sequential freeze-in}.

In this paper we have systematically studied the different regions of the parameter space that generate the observed DM abundance  in the early Universe, within 
a warped extra-dimensional model . We assume that both the SM and the DM particles are localized in the IR-brane, where the effective four-dimensional Planck scale is given by 
$\Lambda$, which is allowed to vary in a wide phenomenological range, $[10^2,\,10^{16}]$~GeV, relaxing the requirement for the RS model to solve the hierarchy problem.
We also include the radion, using the Goldberger-Wise mechanism~\cite{Goldberger:1999uk} to generate the required potential to stabilize the size of the extra dimension.
For definiteness, we consider scalar DM and focus on its interactions with gravitational  mediators, i.e., the radion, the graviton and the KK-gravitons.

As the interaction rates between the visible and the dark sectors have a strong temperature dependence, the bulk of the DM density is typically produced at the highest temperatures 
reached by the SM thermal bath, which in the approximation of a sudden decay of the inflaton corresponds to the reheating temperature, $\Trh$.
This is a characteristic of the so-called UV freeze-in.
We found however a case where the DM abundance was mainly produced at much lower temperatures, corresponding to the sequential freeze-in where the radion was generated via inverse decays.
In that case the peak of the production happens when the temperature approaches the radion mass, $T\simeq m_r/2.5$.

The possibility of generating the DM relic density within the RS scenario via the usual freeze-out mechanism was analyzed in Refs.~\cite{Lee:2013bua, Lee:2014caa, Han:2015cty, Rueter:2017nbk, Rizzo:2018ntg, Carrillo-Monteverde:2018phy, Rizzo:2018joy, Kumar:2019iqs}. After including the DM annihilation channel into KK-gravitons previously disregarded,
it was found that even when both SM and DM particles live in the IR-brane there is a region compatible with the experimental and theoretical constraints where it is possible to reach the correct DM  relic abundance~\cite{Folgado:2019sgz}.
The allowed region corresponds to $m_\chi \in [1,\,15]$~TeV and $\Lambda \in [10,\,10^4]$~TeV.  The upper limit on the DM mass comes from unitarity, while the lower limit 
is an indirect one, derived from searches at LHC of KK-graviton resonant production, which constrains the scale $\Lambda$ as a function of the first KK-graviton mass.
This bound is very relevant, since it determines the minimum value of the DM mass for which the annihilation channel into the first KK-graviton mode is kinematically open, 
leading to the observed DM relic density.
In the freeze-out scenario, the LHC prospects for the near future exclude most part of the allowed region.

In  the present work we find that it is also possible to obtain the correct DM relic abundance in the same RS model via  the freeze-in mechanism, for DM masses in a much wider range spanning typically from the keV to the PeV scale, and  larger values of the scale $\Lambda$ than in the freeze-out scenario. 
This implies that the LHC bounds on the parameter space of the model are weaker than in the freeze-out case. This can be seen in Fig.~\ref{fig:all2}, 
where we summarize our results in the ($m_1$, $\Lambda$) plane for the benchmark DM mass $m_\chi=1$~MeV, finding that only the lower-left corner will be probed by HL-LHC.
 On the other hand, other constraints are relevant, such as the life-time of the radion, which we require to be larger than 1 s to avoid problems with BBN, and excludes the upper-left  corner. 
The results are not strongly dependent of the radion mass: for this reason we fix $m_r = m_{1}/10^3$, in agreement with the expectation  within the Goldberger-Wise mechanism. 
We find that the observed DM  relic density can be obtained in a wide range of reheating temperatures, $\Trh \in [1, 10^{10} ]$~GeV. Notice that we find some region of the parameter
space for which the observed DM relic abundance is achieved with $\Lambda$ as low as a few TeV (with lower values excluded by LHC data). In this region, 
the hierarchy problem is mostly solved, leaving only a remnant little hierarchy to be explained.

Finally, we argued that a more detailed analysis of the present model will require to go beyond the usual approximation where the inflaton decays instantaneously, and therefore the reheat temperature is the maximal temperature reached by the SM thermal bath.
This is due to the strong temperature dependence of some interaction rate densities that enter in the determination of the DM relic abundance.
A complete analysis must take into account the details of the inflationary model (in particular on the energy density carried by the inflaton and its equation of state parameter previous to its decay), and is therefore beyond the scope of this study.

\acknowledgments
We thank Florian Nortier for valuable discussions.
NB thanks the theoretical physics department of University of Valencia and the IFIC for their warm hospitality.
This project has received funding from the European Union's Horizon 2020 research and innovation programme under the Marie Sk\l{}odowska-Curie grant agreements 674896 and 690575.
This work is also supported by the Spanish MINECO Grants SEV-2014-0398 and FPA2017-84543-P, and by Generalitat Valenciana through the ``plan GenT'' program (CIDEGENT/2018/019) and the grant PROMETEO/2019/083.
NB is partially supported by Universidad Antonio Nariño grants 2018204, 2019101 and 2019248.
This research made use of IPython~\cite{Perez:2007emg}, Matplotlib~\cite{Hunter:2007ouj} and SciPy~\cite{SciPy}.


\appendix
\section{Kaluza-Klein decomposition in the Randall-Sundrum scenario}
\label{app:KKdec}

Any 5-dimensional field $\phi_{\mu\nu}$ can be written as a KK tower of 4-dimensional fields as follows:
\be
\phi_{\mu \nu}(x,y) = \sum \phi^{n}_{\mu \nu}(x) \frac{\chi^{n}(y)}{\sqrt{r_c}} \, ,
\ee
being $\chi^{n}(y)$ the wave-functions of the KK-modes along the extra-dimension. 

The equation of motion for the $n^\text{th}$ KK-mode is given by:
\be
\left(\eta^{\mu \nu}\partial_\mu \partial_\nu + m_n^2\right)\, \phi^{n}_{\mu \nu}(x) = 0 \, ,
\ee
where $m_n$ is its mass. Using the Einstein equations we obtain~\cite{Davoudiasl:1999jd}:
\be
-\frac{1}{r_c^2} \frac{d}{dy} \left( e^{-4\sigma} \frac{d\chi^{n}}{dy} \right) = m_n^2\, e^{-2\sigma}\chi^{n} \, ,
\ee
from which:
\be
\chi^{n}(y)=\frac{e^{2\sigma(y)}}{N_n}\left[ J_2(z_n) + \alpha_n Y_2(z_n)\right] \, ,
\ee
being $J_2$ and $Y_2$ Bessel functions of order 2 and $z_n (y) = m_n/k e^{\sigma(y)}$.
The $N_n$ factor is the $n^\text{th}$ KK-mode wave-function normalization.
In the limit $m_n / k \ll 1$ and $e^{k \pi r_c} \gg 1$, the coefficient $\alpha_n$ becomes $\alpha_n \simeq x_n^2 \exp \left (-2 k \,\pi\, r_c \right )$, 
where $x_n$ are the are the roots of the Bessel function, $J_1 (x_n) = 0$, and the masses of the KK-modes are given by: 
\be
m_n = k\, x_n\, e^{- k\, \pi\, r_c} \, .
\ee
Notice that, for low $n$, the KK-modes masses are not equally spaced, as they are proportional to the roots of the Bessel function $J_1$. At large values of $n$, on the other
hand, the roots of the Bessel function become approximately $x_n = \pi \left (n + \frac{1}{4} \right ) + {\cal O} \left ( n^{-1} \right )$. In this limit, the KK-modes masses are
approximately equally spaced (as in LED and the CW/LD scenarios) and proportional to a characteristic length scale $R$ such that:
\be
\label{eq:KKmasseslargen}
m_n \simeq \left ( k\,\pi \, e^{- k\, \pi\, r_c} \right ) \, n = \frac{n}{R} \, ,   \qquad {\left ( n \gg 1 \right) }
\ee
where $R = x_1/m_1 = 1/(k \pi) e^{k \, \pi \, r_c}$ (with $x_1 = 3.81$) is ${\cal O}$ (TeV$^{-1}$).

The normalization factors can be computed imposing that: 
\be
\int dy\, e^{-2\sigma} \left [\chi^{n} \right]^2 = 1 \, .
\ee
In the same approximation as above, i.e. for $m_n / k \ll 1$ and $e^{k \,\pi\, r_c} \gg 1$, we get: 
\begin{equation}
	N_0 = - \frac{1}{\sqrt{k r_c}} \qquad \text{ and } \qquad N_n = \frac{1}{\sqrt{2 k\, r_c}}  \, e^{k\, \pi\, r_c} \, J_2 (x_n) \, .
\end{equation}
Notice the difference between the $n=0$ mode and the $n>0$ modes:  for $n=0$, the wave-function at the IR-brane location $y=\pi$ takes the form
\be
\chi^{0}(y=\pi) = \sqrt{k\, r_c} \left (1-e^{-2 k \,\pi\, r_c} \right ) = -\sqrt{r_c} \, \frac{M_5^{3/2}}{M_P} \, ,
\ee
whereas for $n>0$:
\be
\chi^{n}(y=\pi) = \sqrt{k\, r_c} \, e^{k\, \pi\, r_c} = \sqrt{r_c} \, e^{k\, \pi\, r_c} \frac{M_5^{3/2}}{M_P} = \sqrt{r_c} \, \frac{M_5^{3/2}}{\Lambda}. 
\ee

\section{Radion Lagrangian}
\label{app:radlag}

As it was already reported in the main text, the radion lagrangian is \cite{Goldberger:1999un, Csaki:1999mp}:
\be
\mathcal{L}_r = \frac{1}{2}(\partial_\mu r)(\partial^\mu r) - \frac{1}{2} m_r^2 r^2 +  \frac{1}{\sqrt{6}\Lambda} r T+ \frac{\alpha_\text{EM} \, C_\text{EM}}{8\pi\sqrt{6}\Lambda} r F_{\mu\nu} F^{\mu\nu} + \frac{\alpha_{S}C_{3}}{8 \pi \sqrt{6} \Lambda} r \sum_a F^a_{\mu\nu} F^{a\mu\nu}\, ,
\ee
where $F_{\mu\nu}$, $F^a_{\mu\nu}$ are the Maxwell and $SU(3)_c$ Yang-Mills tensors, respectively.
On the other hand, $C_3$ and $C_\text{EM}$ encode all information about the massless gauge boson contributions and are given by:
\bea
C_3 &=& b_\text{IR}^{(3)} - b_\text{UV}^{(3)} + \frac{1}{2}\sum_q F_{1/2}(x_q) \, , \label{eq:C3}\\
C_\text{EM} &=& b_\text{IR}^\text{(EM)} - b_\text{UV}^\text{(EM)} + F_1(x_W)  - \sum_q N_cQ_{q}^2F_{1/2}(x_q) \, ,\label{eq:CEM}
\eea
with $x_q = 4m_q/m_r$ and $x_W = 4m_w/m_r$. The explicit form of $F_{1/2}$  and the values of the one-loop $\beta$-function coefficients $b$ are given by \cite{Blum:2014jca}:
\bea
F_{1/2}(x) &=& 2x[1 + (1-x)f(x)] , \\
F_{1}(x) &=& 2 + 3x + 3x(2-x)f(x) , 
\eea
\be
f(x) = \left \{
\begin{array}{lll}
	[\arcsin(1/\sqrt{x})]^2 \hphantom{} \hphantom{} \hphantom{} \hphantom{} \hphantom{} &\qquad \text{ for } x>1 , \\
&&\\
-\frac{1}{4}\left[\log\left( \frac{1 + \sqrt{x-1}}{1 - \sqrt{x-1}} \right) - i\pi \right]^2 &\qquad \text{ for } x<1 ,
\end{array}
\right .
\ee
while $b_\text{IR}^\text{(EM)} - b_\text{UV}^\text{(EM)} = 11/3$ and $b_\text{IR}^{(3)} - b_\text{UV}^{(3)} = -11 + 2n/3$, where $n$ is the number of quarks whose mass is smaller than $m_r/2$.

\section{Relevant Interaction Rates}
\label{app:annihil}

In this appendix we report the different cross sections and decay widths used in this analysis, for the case of {\it real scalar} DM.
All relevant Feynman rules can be found in Ref.~\cite{Folgado:2019sgz}.

\subsection{Dark Matter Annihilation}\label{app:DMannihila}
In order to analyze the phenomenology of the FIMP DM in the RS model it is necessary to obtain the interaction rates of DM annihilating into SM particles via the $s$-channel exchange of KK-gravitons or a radion.

\subsubsection{Through KK-gravitons}
\label{app:scalarannihil}

Here we show the different annihilation cross sections of DM $\chi$ into SM particles, mediated by the exchange of KK-gravitons.
In the following expressions we use the notation $S$, $\psi$, $V$ and $v$  for SM scalars, fermions, massive vectors and massless vectors, respectively:
\bea
\sigma(\chi\chi \rightarrow SS) &=& |S_\text{KK}|^2 \frac{s^3}{5760 \, \pi \Lambda^4}\left (1-4\frac{m_{\chi}^2}{s}\right )^\frac32 \left ( 1-4\frac{m_S^2}{s}\right )^\frac52, \\
\sigma(\chi\chi \rightarrow \bar{\psi} \psi) &=& |S_\text{KK}|^2 \frac{s^3}{2880 \, \pi \Lambda^4}\left ( 1 - 4\frac{m_{\psi}^2}{s} \right )^\frac32 \left ( 1 - 4\frac{m_{\chi}^2}{s} \right )^\frac32 
\left (  3 + 8 \frac{m_{\psi}^2}{s} \right ), \\
\sigma(\chi \chi \rightarrow V V) &=&  |S_\text{KK}|^2 \frac{s^3}{5760 \, \pi \Lambda^4}\left (1 - 4\frac{m_{\chi}^2}{s}\right )^\frac32 \left ( 1 - 4\frac{m_V^2}{s} \right )^\frac12 
\left ( 13 + \frac{56 m_V^2}{s} + \frac{48 m_V^4}{s^2} \right ),\qquad\\
\sigma(\chi \chi \rightarrow v v) &=& |S_\text{KK}|^2 \frac{s^3}{480 \, \pi \Lambda^4}\left ( 1-4\frac{m_{\chi}^2}{s}\right )^\frac32,
\eea
where $S_\text{KK}$ corresponds to the sum over all KK-graviton propagators:
\be
S_\text{KK} \equiv \sum_{n=1}^{\infty} \frac{1}{s-m_n^2 + i\,m_n\,\Gamma_n}\,.
\ee

\subsubsection{Through a Radion}
\label{app:scalarannihilrad}

The KK-gravitons are not the only 5-dimensional fields in the bulk.
In fact, in order to stabilize the size of the extra-dimension it is necessary to introduce a new scalar field that mixes with the graviscalar.
The zero-mode of the KK-tower of this new field receives the name of radion and can mediate the DM annihilations into SM states.
The corresponding cross sections are given by: 
\bea
\sigma(\chi \chi \rightarrow SS) &=& \mathcal{P} \frac{s^3}{1152 \, \pi \Lambda^4} \sqrt{\frac{s-4m_S^2}{s-4m_\chi^2}} \left (1+2\frac{m_{\chi}^2}{s}\right )^{2} \left (1+2\frac{m_S^2}{s}\right )^{2} ,  \\
\sigma(\chi \chi \rightarrow \bar{\psi} \psi) &=& \mathcal{P} \frac{s^2 m_{\psi}^2}{288 \, \pi \Lambda^4}\left ( 1 - 4\frac{m_{\psi}^2}{s} \right )^\frac32 \left ( 1 + 2 \frac{m_{\chi}^2}{s} \right )^2 
\left ( 1 - 4\frac{m_{\chi}^2}{s} \right )^{-\frac12}  , \\
\sigma(\chi \chi \rightarrow V V) &=&   \mathcal{P} \frac{s^3}{1152 \, \pi \Lambda^4}\sqrt{\frac{s-4m_V^2}{s-4m_\chi^2}} \left (  1 - 4\frac{m_V^2}{s} + 12\frac{ m_V^4}{s^2}   \right )  ,  \\
\sigma(\chi \chi \rightarrow v v) &=&  \mathcal{P} \frac{s^3\, \alpha_i^2\, C_i^2}{9216 \, \pi^3 \Lambda^4}\left (1+2\frac{m_{\chi}^2}{s}\right )^{2} \left (  1 - 4\frac{m_{\chi}^2}{s} \right )^{-\frac12}   , 
\eea
where $ \mathcal{P} \equiv \left[(s-m_{r}^2)^2 + m_r^2 \,\Gamma_r^2\right]^{-1}$ is the radion propagator.
For the SM massless vectors the vertex is generated by the trace anomaly and, therefore, the cross sections are proportional to $\alpha_\text{EM}$ and $C_\text{EM}$ for the photon case, and to $\alpha_{3}$ and $C_{3}$ for the gluon case, as given in eqs.~\eqref{eq:C3} and~\eqref{eq:CEM}.

\subsection{KK-graviton Annihilation}
\label{app:gravitonanh}

For the sequential freeze-in we are interested in processes that involve KK-graviton $G_n$ annihilations into SM particles.
The corresponding cross-sections can be approximated by:
\bea
	\sigma(G_n  G_n \rightarrow S  S) &\simeq & \frac{1}{96000 \, \pi}\frac{s^5}{\Lambda^4 m_n^8} \, , \\
	\sigma(G_n  G_n \rightarrow \bar{\psi}  \psi ) & \simeq & \frac{1}{604800 \, \pi}\frac{s^5}{\Lambda^4 m_n^8}  \, , \\
	\sigma(G_n  G_n \rightarrow V  V ) & \simeq & \sigma(G_n  G_n  \rightarrow v v)  \simeq   \frac{19}{28800 \, \pi}\frac{s^5}{\Lambda^4 m_n^8} \,.
\eea
Therefore, the total annihilation cross section for the $n^\text{th}$ KK-graviton into SM states becomes:
\begin{equation}\label{CSKKSM}
        \sigma_\text{KK$\to$SM}(s) \simeq \frac{8\times 10^{-3}}{\pi} \frac{s^5}{\Lambda^4 m_n^8}\,.
\end{equation}

\subsection{Radion Annihilation}
\label{app:radionanh}

A second contribution to sequential freeze in comes from the annihilation of a pair of radions into SM particles, and is given by:
\bea
\sigma(rr \rightarrow S  S) &\simeq & \frac{1}{540 \, \pi}\frac{s}{\Lambda^4} \, ,  \\
\sigma(rr \rightarrow \bar{\psi}  \psi) & \simeq & \frac{25}{64 \, \pi}\frac{m_{\psi}^2}{\Lambda^4}  \, , \\
\sigma(rr \rightarrow V V) & \simeq &  \frac{1}{1152 \,  \pi}\frac{s}{\Lambda^4} \, , \\
\sigma(rr \rightarrow v v) & = & 0  \, .
\eea
The total annihilation cross section of radions into SM states becomes:
\begin{equation}\label{CSradionSM}
        \sigma_\text{r$\to$SM}(s) \simeq \frac{9\times 10^{-3}}{\pi} \frac{s}{\Lambda^4} \, ,
\end{equation} 
where the contribution of SM fermions is highly suppressed by their masses and was therefore neglected.
Notice that eqs.~\eqref{CSKKSM} and~\eqref{CSradionSM} do not scale in the same way with the center-of-mass energy $s$, due to their different dependence on the masses.
In particular, the $m^{-8}$ factor in eq.~\eqref{CSKKSM} comes from the polarization tensor of the KK-gravitons (spin-2 massive particles) and is not present in the case of radions (spin-0).

\subsection{KK-graviton Decays}

KK-gravitons can decay into both SM and DM particles. The corresponding decay widths are:
\begin{eqnarray}
	\Gamma_{\text{KK}\to\text{SM}} &\simeq& \frac{73}{240\, \pi} \frac{m_n^3}{\Lambda^2} \, ,\label{eq:decKKtoSM}\\
	\Gamma_{\text{KK}\to\text{DM}} &=& \frac{m_n^3}{960 \, \pi \Lambda^2} \left ( 1 - 4\frac{m_\chi^2}{m_n^2} \right )^{5/2},\label{eq:decKKtoDM}
\end{eqnarray}
where all SM masses were neglected for simplicity.

\subsection{Radion Decays}

Eventually, the decay widths of radions into SM and DM particles are:
\begin{eqnarray}
	\Gamma_{\text{r}\to\text{SM}} &\simeq& \frac{37 m_r^3}{192 \, \pi \Lambda^2} \,,\label{eq:decrtoSM}\\
	\Gamma_{\text{r}\to\text{DM}} &=& \frac{m_r^3}{192 \, \pi \Lambda^2} \left ( 1 - 4\frac{m_\chi^2}{m_r^2} \right )^\frac12 \left (  1 +2\frac{m_\chi^2}{m_r^2} \right )^{2}   \, ,
\end{eqnarray}
where again all SM masses were neglected for simplicity.

\bibliography{biblio}

\providecommand{\href}[2]{#2}\begingroup\raggedright\begin{thebibliography}{10}

\bibitem{Garny:2015sjg}
M.~Garny, M.~Sandora and M.~S. Sloth, \emph{{Planckian Interacting Massive
  Particles as Dark Matter}},
  \href{https://doi.org/10.1103/PhysRevLett.116.101302}{\emph{Phys. Rev. Lett.}
  {\bfseries 116} (2016) 101302},
  [\href{https://arxiv.org/abs/1511.03278}{{\ttfamily 1511.03278}}].

\bibitem{Tang:2017hvq}
Y.~Tang and Y.-L. Wu, \emph{{On Thermal Gravitational Contribution to Particle
  Production and Dark Matter}},
  \href{https://doi.org/10.1016/j.physletb.2017.10.034}{\emph{Phys. Lett. B}
  {\bfseries 774} (2017) 676--681},
  [\href{https://arxiv.org/abs/1708.05138}{{\ttfamily 1708.05138}}].

\bibitem{Garny:2017kha}
M.~Garny, A.~Palessandro, M.~Sandora and M.~S. Sloth, \emph{{Theory and
  Phenomenology of Planckian Interacting Massive Particles as Dark Matter}},
  \href{https://doi.org/10.1088/1475-7516/2018/02/027}{\emph{JCAP} {\bfseries
  02} (2018) 027}, [\href{https://arxiv.org/abs/1709.09688}{{\ttfamily
  1709.09688}}].

\bibitem{Bernal:2018qlk}
N.~Bernal, M.~Dutra, Y.~Mambrini, K.~Olive, M.~Peloso and M.~Pierre,
  \emph{{Spin-2 Portal Dark Matter}},
  \href{https://doi.org/10.1103/PhysRevD.97.115020}{\emph{Phys. Rev.}
  {\bfseries D97} (2018) 115020},
  [\href{https://arxiv.org/abs/1803.01866}{{\ttfamily 1803.01866}}].

\bibitem{Antoniadis:1990ew}
I.~Antoniadis, \emph{{A Possible new dimension at a few TeV}},
  \href{https://doi.org/10.1016/0370-2693(90)90617-F}{\emph{Phys. Lett.}
  {\bfseries B246} (1990) 377--384}.

\bibitem{Antoniadis:1997zg}
I.~Antoniadis, S.~Dimopoulos and G.~Dvali, \emph{{Millimeter range forces in
  superstring theories with weak scale compactification}},
  \href{https://doi.org/10.1016/S0550-3213(97)00808-0}{\emph{Nucl.Phys.}
  {\bfseries B516} (1998) 70--82},
  [\href{https://arxiv.org/abs/hep-ph/9710204}{{\ttfamily hep-ph/9710204}}].

\bibitem{ArkaniHamed:1998rs}
N.~Arkani-Hamed, S.~Dimopoulos and G.~Dvali, \emph{{The Hierarchy problem and
  new dimensions at a millimeter}},
  \href{https://doi.org/10.1016/S0370-2693(98)00466-3}{\emph{Phys.Lett.}
  {\bfseries B429} (1998) 263--272},
  [\href{https://arxiv.org/abs/hep-ph/9803315}{{\ttfamily hep-ph/9803315}}].

\bibitem{Antoniadis:1998ig}
I.~Antoniadis, N.~Arkani-Hamed, S.~Dimopoulos and G.~Dvali, \emph{{New
  dimensions at a millimeter to a Fermi and superstrings at a TeV}},
  \href{https://doi.org/10.1016/S0370-2693(98)00860-0}{\emph{Phys.Lett.}
  {\bfseries B436} (1998) 257--263},
  [\href{https://arxiv.org/abs/hep-ph/9804398}{{\ttfamily hep-ph/9804398}}].

\bibitem{ArkaniHamed:1998nn}
N.~Arkani-Hamed, S.~Dimopoulos and G.~Dvali, \emph{{Phenomenology, astrophysics
  and cosmology of theories with submillimeter dimensions and TeV scale quantum
  gravity}}, \href{https://doi.org/10.1103/PhysRevD.59.086004}{\emph{Phys.Rev.}
  {\bfseries D59} (1999) 086004},
  [\href{https://arxiv.org/abs/hep-ph/9807344}{{\ttfamily hep-ph/9807344}}].

\bibitem{Randall:1999ee}
L.~Randall and R.~Sundrum, \emph{{A Large mass hierarchy from a small extra
  dimension}}, \href{https://doi.org/10.1103/PhysRevLett.83.3370}{\emph{Phys.
  Rev. Lett.} {\bfseries 83} (1999) 3370--3373},
  [\href{https://arxiv.org/abs/hep-ph/9905221}{{\ttfamily hep-ph/9905221}}].

\bibitem{Randall:1999vf}
L.~Randall and R.~Sundrum, \emph{{An Alternative to compactification}},
  \href{https://doi.org/10.1103/PhysRevLett.83.4690}{\emph{Phys. Rev. Lett.}
  {\bfseries 83} (1999) 4690--4693},
  [\href{https://arxiv.org/abs/hep-th/9906064}{{\ttfamily hep-th/9906064}}].

\bibitem{Antoniadis:2011qw}
I.~Antoniadis, A.~Arvanitaki, S.~Dimopoulos and A.~Giveon, \emph{{Phenomenology
  of TeV Little String Theory from Holography}},
  \href{https://doi.org/10.1103/PhysRevLett.108.081602}{\emph{Phys. Rev. Lett.}
  {\bfseries 108} (2012) 081602},
  [\href{https://arxiv.org/abs/1102.4043}{{\ttfamily 1102.4043}}].

\bibitem{Cox:2012ee}
P.~Cox and T.~Gherghetta, \emph{{Radion Dynamics and Phenomenology in the
  Linear Dilaton Model}},
  \href{https://doi.org/10.1007/JHEP05(2012)149}{\emph{JHEP} {\bfseries 05}
  (2012) 149}, [\href{https://arxiv.org/abs/1203.5870}{{\ttfamily 1203.5870}}].

\bibitem{Giudice:2016yja}
G.~F. Giudice and M.~McCullough, \emph{{A Clockwork Theory}},
  \href{https://doi.org/10.1007/JHEP02(2017)036}{\emph{JHEP} {\bfseries 02}
  (2017) 036}, [\href{https://arxiv.org/abs/1610.07962}{{\ttfamily
  1610.07962}}].

\bibitem{Giudice:2017fmj}
G.~F. Giudice, Y.~Kats, M.~McCullough, R.~Torre and A.~Urbano,
  \emph{{Clockwork/linear dilaton: structure and phenomenology}},
  \href{https://doi.org/10.1007/JHEP06(2018)009}{\emph{JHEP} {\bfseries 06}
  (2018) 009}, [\href{https://arxiv.org/abs/1711.08437}{{\ttfamily
  1711.08437}}].

\bibitem{Lee:2013bua}
H.~M. Lee, M.~Park and V.~Sanz, \emph{{Gravity-mediated (or Composite) Dark
  Matter}}, \href{https://doi.org/10.1140/epjc/s10052-014-2715-8}{\emph{Eur.
  Phys. J.} {\bfseries C74} (2014) 2715},
  [\href{https://arxiv.org/abs/1306.4107}{{\ttfamily 1306.4107}}].

\bibitem{Lee:2014caa}
H.~M. Lee, M.~Park and V.~Sanz, \emph{{Gravity-mediated (or Composite) Dark
  Matter Confronts Astrophysical Data}},
  \href{https://doi.org/10.1007/JHEP05(2014)063}{\emph{JHEP} {\bfseries 05}
  (2014) 063}, [\href{https://arxiv.org/abs/1401.5301}{{\ttfamily 1401.5301}}].

\bibitem{Han:2015cty}
C.~Han, H.~M. Lee, M.~Park and V.~Sanz, \emph{{The diphoton resonance as a
  gravity mediator of dark matter}},
  \href{https://doi.org/10.1016/j.physletb.2016.02.040}{\emph{Phys. Lett.}
  {\bfseries B755} (2016) 371--379},
  [\href{https://arxiv.org/abs/1512.06376}{{\ttfamily 1512.06376}}].

\bibitem{Rueter:2017nbk}
T.~D. Rueter, T.~G. Rizzo and J.~L. Hewett, \emph{{Gravity-Mediated Dark Matter
  Annihilation in the Randall-Sundrum Model}},
  \href{https://doi.org/10.1007/JHEP10(2017)094}{\emph{JHEP} {\bfseries 10}
  (2017) 094}, [\href{https://arxiv.org/abs/1706.07540}{{\ttfamily
  1706.07540}}].

\bibitem{Rizzo:2018ntg}
T.~G. Rizzo, \emph{{Kinetic mixing, dark photons and an extra dimension. Part
  I}}, \href{https://doi.org/10.1007/JHEP07(2018)118}{\emph{JHEP} {\bfseries
  07} (2018) 118}, [\href{https://arxiv.org/abs/1801.08525}{{\ttfamily
  1801.08525}}].

\bibitem{Carrillo-Monteverde:2018phy}
A.~Carrillo-Monteverde, Y.-J. Kang, H.~M. Lee, M.~Park and V.~Sanz, \emph{{Dark
  Matter Direct Detection from new interactions in models with spin-two
  mediators}}, \href{https://doi.org/10.1007/JHEP06(2018)037}{\emph{JHEP}
  {\bfseries 06} (2018) 037},
  [\href{https://arxiv.org/abs/1803.02144}{{\ttfamily 1803.02144}}].

\bibitem{Rizzo:2018joy}
T.~G. Rizzo, \emph{{Kinetic mixing, dark photons and extra dimensions. Part II:
  fermionic dark matter}},
  \href{https://doi.org/10.1007/JHEP10(2018)069}{\emph{JHEP} {\bfseries 10}
  (2018) 069}, [\href{https://arxiv.org/abs/1805.08150}{{\ttfamily
  1805.08150}}].

\bibitem{Brax:2019koq}
P.~Brax, S.~Fichet and P.~Tanedo, \emph{{The Warped Dark Sector}},
  \href{https://doi.org/10.1016/j.physletb.2019.135012}{\emph{Phys. Lett. B}
  {\bfseries 798} (2019) 135012},
  [\href{https://arxiv.org/abs/1906.02199}{{\ttfamily 1906.02199}}].

\bibitem{Folgado:2019sgz}
M.~G. Folgado, A.~Donini and N.~Rius, \emph{{Gravity-mediated Scalar Dark
  Matter in Warped Extra-Dimensions}},
  \href{https://doi.org/10.1007/JHEP01(2020)161}{\emph{JHEP} {\bfseries 01}
  (2020) 161}, [\href{https://arxiv.org/abs/1907.04340}{{\ttfamily
  1907.04340}}].

\bibitem{Kumar:2019iqs}
M.~Kumar, A.~Goyal and R.~Islam, \emph{{Dark matter in the Randall-Sundrum
  model}},  in \emph{{64$^{th}$ Annual Conference of the South African
  Institute of Physics (SAIP2019) Polokwane, South Africa, July 8-12, 2019}},
  2019, \href{https://arxiv.org/abs/1908.10334}{{\ttfamily 1908.10334}}.

\bibitem{Kang:2020huh}
Y.-J. Kang and H.~M. Lee, \emph{{Lightening Gravity-Mediated Dark Matter}},
  \href{https://doi.org/10.1140/epjc/s10052-020-8153-x}{\emph{Eur. Phys. J. C}
  {\bfseries 80} (2020) 602},
  [\href{https://arxiv.org/abs/2001.04868}{{\ttfamily 2001.04868}}].

\bibitem{Chivukula:2020hvi}
R.~S. Chivukula, D.~Foren, K.~A. Mohan, D.~Sengupta and E.~H. Simmons,
  \emph{{Massive Spin-2 Scattering Amplitudes in Extra-Dimensional Theories}},
  \href{https://doi.org/10.1103/PhysRevD.101.075013}{\emph{Phys. Rev. D}
  {\bfseries 101} (2020) 075013},
  [\href{https://arxiv.org/abs/2002.12458}{{\ttfamily 2002.12458}}].

\bibitem{Kang:2020yul}
Y.-J. Kang and H.~M. Lee, \emph{{Dark matter self-interactions from spin-2
  mediators}},  \href{https://arxiv.org/abs/2002.12779}{{\ttfamily
  2002.12779}}.

\bibitem{Kang:2020afi}
Y.-J. Kang and H.~M. Lee, \emph{{Effective theory for self-interacting dark
  matter and massive spin-2 mediators}},
  \href{https://arxiv.org/abs/2003.09290}{{\ttfamily 2003.09290}}.

\bibitem{Folgado:2019gie}
M.~G. Folgado, A.~Donini and N.~Rius, \emph{{Gravity-mediated Dark Matter in
  Clockwork/Linear Dilaton Extra-Dimensions}},
  \href{https://doi.org/10.1007/JHEP04(2020)036}{\emph{JHEP} {\bfseries 04}
  (2020) 036}, [\href{https://arxiv.org/abs/1912.02689}{{\ttfamily
  1912.02689}}].

\bibitem{McDonald:2001vt}
J.~McDonald, \emph{{Thermally generated gauge singlet scalars as
  selfinteracting dark matter}},
  \href{https://doi.org/10.1103/PhysRevLett.88.091304}{\emph{Phys.Rev.Lett.}
  {\bfseries 88} (2002) 091304},
  [\href{https://arxiv.org/abs/hep-ph/0106249}{{\ttfamily hep-ph/0106249}}].

\bibitem{Choi:2005vq}
K.-Y. Choi and L.~Roszkowski, \emph{{E-WIMPs}},
  \href{https://doi.org/10.1063/1.2149672}{\emph{AIP Conf. Proc.} {\bfseries
  805} (2006) 30--36}, [\href{https://arxiv.org/abs/hep-ph/0511003}{{\ttfamily
  hep-ph/0511003}}].

\bibitem{Kusenko:2006rh}
A.~Kusenko, \emph{{Sterile neutrinos, dark matter, and the pulsar velocities in
  models with a Higgs singlet}},
  \href{https://doi.org/10.1103/PhysRevLett.97.241301}{\emph{Phys. Rev. Lett.}
  {\bfseries 97} (2006) 241301},
  [\href{https://arxiv.org/abs/hep-ph/0609081}{{\ttfamily hep-ph/0609081}}].

\bibitem{Petraki:2007gq}
K.~Petraki and A.~Kusenko, \emph{{Dark-matter sterile neutrinos in models with
  a gauge singlet in the Higgs sector}},
  \href{https://doi.org/10.1103/PhysRevD.77.065014}{\emph{Phys. Rev.}
  {\bfseries D77} (2008) 065014},
  [\href{https://arxiv.org/abs/0711.4646}{{\ttfamily 0711.4646}}].

\bibitem{Hall:2009bx}
L.~J. Hall, K.~Jedamzik, J.~March-Russell and S.~M. West, \emph{{Freeze-In
  Production of FIMP Dark Matter}},
  \href{https://doi.org/10.1007/JHEP03(2010)080}{\emph{JHEP} {\bfseries 1003}
  (2010) 080}, [\href{https://arxiv.org/abs/0911.1120}{{\ttfamily 0911.1120}}].

\bibitem{Bernal:2017kxu}
N.~Bernal, M.~Heikinheimo, T.~Tenkanen, K.~Tuominen and V.~Vaskonen, \emph{{The
  Dawn of FIMP Dark Matter: A Review of Models and Constraints}},
  \href{https://doi.org/10.1142/S0217751X1730023X}{\emph{Int. J. Mod. Phys.}
  {\bfseries A32} (2017) 1730023},
  [\href{https://arxiv.org/abs/1706.07442}{{\ttfamily 1706.07442}}].

\bibitem{Elahi:2014fsa}
F.~Elahi, C.~Kolda and J.~Unwin, \emph{{UltraViolet Freeze-in}},
  \href{https://doi.org/10.1007/JHEP03(2015)048}{\emph{JHEP} {\bfseries 03}
  (2015) 048}, [\href{https://arxiv.org/abs/1410.6157}{{\ttfamily 1410.6157}}].

\bibitem{Goldberger:1999uk}
W.~D. Goldberger and M.~B. Wise, \emph{{Modulus stabilization with bulk
  fields}}, \href{https://doi.org/10.1103/PhysRevLett.83.4922}{\emph{Phys. Rev.
  Lett.} {\bfseries 83} (1999) 4922--4925},
  [\href{https://arxiv.org/abs/hep-ph/9907447}{{\ttfamily hep-ph/9907447}}].

\bibitem{Sarkar:1995dd}
S.~Sarkar, \emph{{Big bang nucleosynthesis and physics beyond the standard
  model}}, \href{https://doi.org/10.1088/0034-4885/59/12/001}{\emph{Rept. Prog.
  Phys.} {\bfseries 59} (1996) 1493--1610},
  [\href{https://arxiv.org/abs/hep-ph/9602260}{{\ttfamily hep-ph/9602260}}].

\bibitem{Kawasaki:2000en}
M.~Kawasaki, K.~Kohri and N.~Sugiyama, \emph{{MeV scale reheating temperature
  and thermalization of neutrino background}},
  \href{https://doi.org/10.1103/PhysRevD.62.023506}{\emph{Phys. Rev.}
  {\bfseries D62} (2000) 023506},
  [\href{https://arxiv.org/abs/astro-ph/0002127}{{\ttfamily
  astro-ph/0002127}}].

\bibitem{Hannestad:2004px}
S.~Hannestad, \emph{{What is the lowest possible reheating temperature?}},
  \href{https://doi.org/10.1103/PhysRevD.70.043506}{\emph{Phys. Rev.}
  {\bfseries D70} (2004) 043506},
  [\href{https://arxiv.org/abs/astro-ph/0403291}{{\ttfamily
  astro-ph/0403291}}].

\bibitem{DeBernardis:2008zz}
F.~De~Bernardis, L.~Pagano and A.~Melchiorri, \emph{{New constraints on the
  reheating temperature of the universe after WMAP-5}},
  \href{https://doi.org/10.1016/j.astropartphys.2008.09.005}{\emph{Astropart.
  Phys.} {\bfseries 30} (2008) 192--195}.

\bibitem{deSalas:2015glj}
P.~F. de~Salas, M.~Lattanzi, G.~Mangano, G.~Miele, S.~Pastor and O.~Pisanti,
  \emph{{Bounds on very low reheating scenarios after Planck}},
  \href{https://doi.org/10.1103/PhysRevD.92.123534}{\emph{Phys. Rev.}
  {\bfseries D92} (2015) 123534},
  [\href{https://arxiv.org/abs/1511.00672}{{\ttfamily 1511.00672}}].

\bibitem{Hasegawa:2019jsa}
T.~Hasegawa, N.~Hiroshima, K.~Kohri, R.~S. Hansen, T.~Tram and S.~Hannestad,
  \emph{{MeV-scale reheating temperature and thermalization of oscillating
  neutrinos by radiative and hadronic decays of massive particles}},
  \href{https://doi.org/10.1088/1475-7516/2019/12/012}{\emph{JCAP} {\bfseries
  12} (2019) 012}, [\href{https://arxiv.org/abs/1908.10189}{{\ttfamily
  1908.10189}}].

\bibitem{Appelquist:1982zs}
T.~Appelquist and A.~Chodos, \emph{{Quantum Effects in Kaluza-Klein Theories}},
  \href{https://doi.org/10.1103/PhysRevLett.50.141}{\emph{Phys. Rev. Lett.}
  {\bfseries 50} (1983) 141}.

\bibitem{Appelquist:1983vs}
T.~Appelquist and A.~Chodos, \emph{{The Quantum Dynamics of Kaluza-Klein
  Theories}}, \href{https://doi.org/10.1103/PhysRevD.28.772}{\emph{Phys. Rev.}
  {\bfseries D28} (1983) 772}.

\bibitem{deWit:1988xki}
B.~de~Wit, M.~Luscher and H.~Nicolai, \emph{{The Supermembrane Is Unstable}},
  \href{https://doi.org/10.1016/0550-3213(89)90214-9}{\emph{Nucl. Phys.}
  {\bfseries B320} (1989) 135--159}.

\bibitem{Ponton:2001hq}
E.~Pontón and E.~Poppitz, \emph{{Casimir energy and radius stabilization in
  five-dimensional orbifolds and six-dimensional orbifolds}},
  \href{https://doi.org/10.1088/1126-6708/2001/06/019}{\emph{JHEP} {\bfseries
  06} (2001) 019}, [\href{https://arxiv.org/abs/hep-ph/0105021}{{\ttfamily
  hep-ph/0105021}}].

\bibitem{Goldberger:1999wh}
W.~D. Goldberger and M.~B. Wise, \emph{{Bulk fields in the Randall-Sundrum
  compactification scenario}},
  \href{https://doi.org/10.1103/PhysRevD.60.107505}{\emph{Phys. Rev.}
  {\bfseries D60} (1999) 107505},
  [\href{https://arxiv.org/abs/hep-ph/9907218}{{\ttfamily hep-ph/9907218}}].

\bibitem{Goldberger:1999un}
W.~D. Goldberger and M.~B. Wise, \emph{{Phenomenology of a stabilized
  modulus}}, \href{https://doi.org/10.1016/S0370-2693(00)00099-X}{\emph{Phys.
  Lett.} {\bfseries B475} (2000) 275--279},
  [\href{https://arxiv.org/abs/hep-ph/9911457}{{\ttfamily hep-ph/9911457}}].

\bibitem{Blum:2014jca}
K.~Blum, M.~Cliche, C.~Csáki and S.~J. Lee, \emph{{WIMP Dark Matter through
  the Dilaton Portal}},
  \href{https://doi.org/10.1007/JHEP03(2015)099}{\emph{JHEP} {\bfseries 03}
  (2015) 099}, [\href{https://arxiv.org/abs/1410.1873}{{\ttfamily 1410.1873}}].

\bibitem{Csaki:1999mp}
C.~Csáki, M.~Graesser, L.~Randall and J.~Terning, \emph{{Cosmology of brane
  models with radion stabilization}},
  \href{https://doi.org/10.1103/PhysRevD.62.045015}{\emph{Phys. Rev.}
  {\bfseries D62} (2000) 045015},
  [\href{https://arxiv.org/abs/hep-ph/9911406}{{\ttfamily hep-ph/9911406}}].

\bibitem{Aharony:1999ti}
O.~Aharony, S.~S. Gubser, J.~M. Maldacena, H.~Ooguri and Y.~Oz, \emph{{Large N
  field theories, string theory and gravity}},
  \href{https://doi.org/10.1016/S0370-1573(99)00083-6}{\emph{Phys. Rept.}
  {\bfseries 323} (2000) 183--386},
  [\href{https://arxiv.org/abs/hep-th/9905111}{{\ttfamily hep-th/9905111}}].

\bibitem{Drees:2015exa}
M.~Drees, F.~Hajkarim and E.~R. Schmitz, \emph{{The Effects of QCD Equation of
  State on the Relic Density of WIMP Dark Matter}},
  \href{https://doi.org/10.1088/1475-7516/2015/06/025}{\emph{JCAP} {\bfseries
  1506} (2015) 025}, [\href{https://arxiv.org/abs/1503.03513}{{\ttfamily
  1503.03513}}].

\bibitem{Hambye:2019dwd}
T.~Hambye, M.~H.~G. Tytgat, J.~Vandecasteele and L.~Vanderheyden, \emph{{Dark
  matter from dark photons: a taxonomy of dark matter production}},
  \href{https://doi.org/10.1103/PhysRevD.100.095018}{\emph{Phys. Rev.}
  {\bfseries D100} (2019) 095018},
  [\href{https://arxiv.org/abs/1908.09864}{{\ttfamily 1908.09864}}].

\bibitem{Bernal:2015xba}
N.~Bernal and X.~Chu, \emph{{$\mathbb{Z}_2$ SIMP Dark Matter}},
  \href{https://doi.org/10.1088/1475-7516/2016/01/006}{\emph{JCAP} {\bfseries
  1601} (2016) 006}, [\href{https://arxiv.org/abs/1510.08527}{{\ttfamily
  1510.08527}}].

\bibitem{Bernal:2017mqb}
N.~Bernal, X.~Chu and J.~Pradler, \emph{{Simply split strongly interacting
  massive particles}},
  \href{https://doi.org/10.1103/PhysRevD.95.115023}{\emph{Phys. Rev.}
  {\bfseries D95} (2017) 115023},
  [\href{https://arxiv.org/abs/1702.04906}{{\ttfamily 1702.04906}}].

\bibitem{Bernal:2020gzm}
N.~Bernal, \emph{{Boosting Freeze-in through Thermalization}},
  \href{https://arxiv.org/abs/2005.08988}{{\ttfamily 2005.08988}}.

\bibitem{Yaguna:2011qn}
C.~E. Yaguna, \emph{{The Singlet Scalar as FIMP Dark Matter}},
  \href{https://doi.org/10.1007/JHEP08(2011)060}{\emph{JHEP} {\bfseries 08}
  (2011) 060}, [\href{https://arxiv.org/abs/1105.1654}{{\ttfamily 1105.1654}}].

\bibitem{Bernal:2018kcw}
N.~Bernal, C.~Cosme, T.~Tenkanen and V.~Vaskonen, \emph{{Scalar singlet dark
  matter in non-standard cosmologies}},
  \href{https://doi.org/10.1140/epjc/s10052-019-6550-9}{\emph{Eur. Phys. J.}
  {\bfseries C79} (2019) 30},
  [\href{https://arxiv.org/abs/1806.11122}{{\ttfamily 1806.11122}}].

\bibitem{Aghanim:2018eyx}
{\scshape Planck} collaboration, N.~Aghanim et~al., \emph{{Planck 2018 results.
  VI. Cosmological parameters}},
  \href{https://arxiv.org/abs/1807.06209}{{\ttfamily 1807.06209}}.

\bibitem{Aaboud:2017buh}
{\scshape ATLAS} collaboration, M.~Aaboud et~al., \emph{{Search for new
  high-mass phenomena in the dilepton final state using 36 fb$^{-1}$ of
  proton-proton collision data at $ \sqrt{s}=13 $ TeV with the ATLAS
  detector}}, \href{https://doi.org/10.1007/JHEP10(2017)182}{\emph{JHEP}
  {\bfseries 10} (2017) 182},
  [\href{https://arxiv.org/abs/1707.02424}{{\ttfamily 1707.02424}}].

\bibitem{Aaboud:2017yyg}
{\scshape ATLAS} collaboration, M.~Aaboud et~al., \emph{{Search for new
  phenomena in high-mass diphoton final states using 37~fb$^{-1}$ of
  proton--proton collisions collected at $\sqrt{s}=13$ TeV with the ATLAS
  detector}}, \href{https://doi.org/10.1016/j.physletb.2017.10.039}{\emph{Phys.
  Lett.} {\bfseries B775} (2017) 105--125},
  [\href{https://arxiv.org/abs/1707.04147}{{\ttfamily 1707.04147}}].

\bibitem{Scherrer:1984fd}
R.~J. Scherrer and M.~S. Turner, \emph{{Decaying Particles Do Not Heat Up the
  Universe}}, \href{https://doi.org/10.1103/PhysRevD.31.681}{\emph{Phys. Rev.}
  {\bfseries D31} (1985) 681}.

\bibitem{Giudice:2000ex}
G.~F. Giudice, E.~W. Kolb and A.~Riotto, \emph{{Largest temperature of the
  radiation era and its cosmological implications}},
  \href{https://doi.org/10.1103/PhysRevD.64.023508}{\emph{Phys. Rev.}
  {\bfseries D64} (2001) 023508},
  [\href{https://arxiv.org/abs/hep-ph/0005123}{{\ttfamily hep-ph/0005123}}].

\bibitem{Garcia:2017tuj}
M.~A.~G. Garcia, Y.~Mambrini, K.~A. Olive and M.~Peloso, \emph{{Enhancement of
  the Dark Matter Abundance Before Reheating: Applications to Gravitino Dark
  Matter}}, \href{https://doi.org/10.1103/PhysRevD.96.103510}{\emph{Phys. Rev.}
  {\bfseries D96} (2017) 103510},
  [\href{https://arxiv.org/abs/1709.01549}{{\ttfamily 1709.01549}}].

\bibitem{Bernal:2019mhf}
N.~Bernal, F.~Elahi, C.~Maldonado and J.~Unwin, \emph{{Ultraviolet Freeze-in
  and Non-Standard Cosmologies}},
  \href{https://doi.org/10.1088/1475-7516/2019/11/026}{\emph{JCAP} {\bfseries
  1911} (2019) 026}, [\href{https://arxiv.org/abs/1909.07992}{{\ttfamily
  1909.07992}}].

\bibitem{Garcia:2020eof}
M.~A. Garcia, K.~Kaneta, Y.~Mambrini and K.~A. Olive, \emph{{Reheating and
  Post-inflationary Production of Dark Matter}},
  \href{https://doi.org/10.1103/PhysRevD.101.123507}{\emph{Phys. Rev. D}
  {\bfseries 101} (2020) 123507},
  [\href{https://arxiv.org/abs/2004.08404}{{\ttfamily 2004.08404}}].

\bibitem{Bernal:2020qyu}
N.~Bernal, J.~Rubio and H.~Veermäe, \emph{{UV Freeze-in in Starobinsky
  Inflation}},  \href{https://arxiv.org/abs/2006.02442}{{\ttfamily
  2006.02442}}.

\bibitem{Chen:2017kvz}
S.-L. Chen and Z.~Kang, \emph{{On UltraViolet Freeze-in Dark Matter during
  Reheating}}, \href{https://doi.org/10.1088/1475-7516/2018/05/036}{\emph{JCAP}
  {\bfseries 1805} (2018) 036},
  [\href{https://arxiv.org/abs/1711.02556}{{\ttfamily 1711.02556}}].

\bibitem{Bhattacharyya:2018evo}
G.~Bhattacharyya, M.~Dutra, Y.~Mambrini and M.~Pierre, \emph{{Freezing-in dark
  matter through a heavy invisible Z'}},
  \href{https://doi.org/10.1103/PhysRevD.98.035038}{\emph{Phys. Rev.}
  {\bfseries D98} (2018) 035038},
  [\href{https://arxiv.org/abs/1806.00016}{{\ttfamily 1806.00016}}].

\bibitem{Chowdhury:2018tzw}
D.~Chowdhury, E.~Dudas, M.~Dutra and Y.~Mambrini, \emph{{Moduli Portal Dark
  Matter}}, \href{https://doi.org/10.1103/PhysRevD.99.095028}{\emph{Phys. Rev.}
  {\bfseries D99} (2019) 095028},
  [\href{https://arxiv.org/abs/1811.01947}{{\ttfamily 1811.01947}}].

\bibitem{Kaneta:2019zgw}
K.~Kaneta, Y.~Mambrini and K.~A. Olive, \emph{{Radiative production of
  nonthermal dark matter}},
  \href{https://doi.org/10.1103/PhysRevD.99.063508}{\emph{Phys. Rev.}
  {\bfseries D99} (2019) 063508},
  [\href{https://arxiv.org/abs/1901.04449}{{\ttfamily 1901.04449}}].

\bibitem{Banerjee:2019asa}
A.~Banerjee, G.~Bhattacharyya, D.~Chowdhury and Y.~Mambrini, \emph{{Dark matter
  seeping through dynamic gauge kinetic mixing}},
  \href{https://doi.org/10.1088/1475-7516/2019/12/009}{\emph{JCAP} {\bfseries
  1912} (2019) 009}, [\href{https://arxiv.org/abs/1905.11407}{{\ttfamily
  1905.11407}}].

\bibitem{Chanda:2019xyl}
P.~Chanda, S.~Hamdan and J.~Unwin, \emph{{Reviving $Z$ and Higgs Mediated Dark
  Matter Models in Matter Dominated Freeze-out}},
  \href{https://doi.org/10.1088/1475-7516/2020/01/034}{\emph{JCAP} {\bfseries
  2001} (2020) 034}, [\href{https://arxiv.org/abs/1911.02616}{{\ttfamily
  1911.02616}}].

\bibitem{Baules:2019zwk}
V.~Baules, N.~Okada and S.~Okada, \emph{{Braneworld Cosmological Effect on
  Freeze-in Dark Matter Density and Lifetime Frontier}},
  \href{https://arxiv.org/abs/1911.05344}{{\ttfamily 1911.05344}}.

\bibitem{Dutra:2019xet}
M.~Dutra, \emph{{Freeze-in production of dark matter through spin-1 and spin-2
  portals}},  in \emph{{PoS(LeptonPhoton2019)076}}, 2019,
  \href{https://arxiv.org/abs/1911.11844}{{\ttfamily 1911.11844}}.

\bibitem{Dutra:2019nhh}
M.~Dutra, \emph{{The moduli portal to dark matter particles}},  in \emph{{11th
  International Symposium on Quantum Theory and Symmetries (QTS2019) Montreal,
  Canada, July 1-5, 2019}}, 2019,
  \href{https://arxiv.org/abs/1911.11862}{{\ttfamily 1911.11862}}.

\bibitem{Mahanta:2019sfo}
D.~Mahanta and D.~Borah, \emph{{TeV Scale Leptogenesis with Dark Matter in
  Non-standard Cosmology}},
  \href{https://doi.org/10.1088/1475-7516/2020/04/032}{\emph{JCAP} {\bfseries
  04} (2020) 032}, [\href{https://arxiv.org/abs/1912.09726}{{\ttfamily
  1912.09726}}].

\bibitem{Cosme:2020mck}
C.~Cosme, M.~Dutra, T.~Ma, Y.~Wu and L.~Yang, \emph{{Neutrino Portal to FIMP
  Dark Matter with an Early Matter Era}},
  \href{https://arxiv.org/abs/2003.01723}{{\ttfamily 2003.01723}}.

\bibitem{Bernal:2020bfj}
N.~Bernal, J.~Rubio and H.~Veermäe, \emph{{Boosting Ultraviolet Freeze-in in
  NO Models}}, \href{https://doi.org/10.1088/1475-7516/2020/06/047}{\emph{JCAP}
  {\bfseries 06} (2020) 047},
  [\href{https://arxiv.org/abs/2004.13706}{{\ttfamily 2004.13706}}].

\bibitem{Perez:2007emg}
F.~P\'erez and B.~E. Granger, \emph{{IPython: A System for Interactive
  Scientific Computing}},
  \href{https://doi.org/10.1109/MCSE.2007.53}{\emph{Comput. Sci. Eng.}
  {\bfseries 9} (2007) 21--29}.

\bibitem{Hunter:2007ouj}
J.~D. Hunter, \emph{{Matplotlib: A 2D Graphics Environment}},
  \href{https://doi.org/10.1109/MCSE.2007.55}{\emph{Comput. Sci. Eng.}
  {\bfseries 9} (2007) 90--95}.

\bibitem{SciPy}
E.~Jones, T.~Oliphant, P.~Peterson et~al., \emph{{SciPy}: Open source
  scientific tools for {Python}},  2001--.

\bibitem{Davoudiasl:1999jd}
H.~Davoudiasl, J.~L. Hewett and T.~G. Rizzo, \emph{{Phenomenology of the
  Randall-Sundrum Gauge Hierarchy Model}},
  \href{https://doi.org/10.1103/PhysRevLett.84.2080}{\emph{Phys. Rev. Lett.}
  {\bfseries 84} (2000) 2080},
  [\href{https://arxiv.org/abs/hep-ph/9909255}{{\ttfamily hep-ph/9909255}}].

\end{thebibliography}\endgroup

\end{document}